\pdfoutput=1

\documentclass[12pt,a4paper]{article}

\usepackage{ifthen} 
\newboolean{pdflatex}
\setboolean{pdflatex}{true} 

\newboolean{articletitles}
\setboolean{articletitles}{true} 

\newboolean{uprightparticles}
\setboolean{uprightparticles}{false} 

\newboolean{inbibliography}
\setboolean{inbibliography}{false} 

\def\paperauthors{LHCb collaboration} 
\def\paperasciititle{Measurement of CP violation in B0->D+pi- decays} 
\def\papertitle{Measurement of \CP violation in \BdDpi decays} 
\def\paperkeywords{{High Energy Physics}, {LHCb}} 
\def\papercopyright{2018 CERN for the benefit of the LHCb Collaboration}
\def\paperlicence{CC-BY-4.0}
\def\paperlicenceurl{https://creativecommons.org/licenses/by/4.0/}

\usepackage[top=1in, bottom=1.25in, left=1in, right=1in]{geometry}

\usepackage{microtype}
\usepackage{lineno}  
\usepackage{xspace} 
\usepackage{caption} 

\usepackage{graphicx}  
\usepackage{color}
\usepackage{colortbl}
\graphicspath{{./figs/}} 

\usepackage{amsmath} 
\usepackage{amssymb}
\usepackage{amsfonts}
\usepackage{upgreek} 

\newcommand*\patchAmsMathEnvironmentForLineno[1]{%
\expandafter\let\csname old#1\expandafter\endcsname\csname #1\endcsname
\expandafter\let\csname oldend#1\expandafter\endcsname\csname
end#1\endcsname
 \renewenvironment{#1}%
   {\linenomath\csname old#1\endcsname}%
   {\csname oldend#1\endcsname\endlinenomath}%
}
\newcommand*\patchBothAmsMathEnvironmentsForLineno[1]{%
  \patchAmsMathEnvironmentForLineno{#1}%
  \patchAmsMathEnvironmentForLineno{#1*}%
}
\AtBeginDocument{%
\patchBothAmsMathEnvironmentsForLineno{equation}%
\patchBothAmsMathEnvironmentsForLineno{align}%
\patchBothAmsMathEnvironmentsForLineno{flalign}%
\patchBothAmsMathEnvironmentsForLineno{alignat}%
\patchBothAmsMathEnvironmentsForLineno{gather}%
\patchBothAmsMathEnvironmentsForLineno{multline}%
\patchBothAmsMathEnvironmentsForLineno{eqnarray}%
}

\usepackage{hyperxmp}

\usepackage[pdftex,
            pdfauthor={\paperauthors},
            pdftitle={\paperasciititle},
            pdfkeywords={\paperkeywords},
            pdfcopyright={Copyright (C) \papercopyright},
            pdflicenseurl={\paperlicenceurl}]{hyperref}

\usepackage[all]{hypcap} 

\usepackage{xspace} 
\usepackage{upgreek}

\def\lhcb {\mbox{LHCb}\xspace}

\def\MagUp {\mbox{\em Mag\kern -0.05em Up}\xspace}

\ifthenelse{\boolean{uprightparticles}}%
{

 \def\Ppi         {\ensuremath{\uppi}\xspace}

 \def\Ppsi        {\ensuremath{\uppsi}\xspace}

 \def\PDelta      {\ensuremath{\Delta}\xspace}                 
 \def\PXi      {\ensuremath{\Xi}\xspace}                 
 \def\PLambda      {\ensuremath{\Lambda}\xspace}                 
 \def\PSigma      {\ensuremath{\Sigma}\xspace}                 
 \def\POmega      {\ensuremath{\Omega}\xspace}                 
 \def\PUpsilon      {\ensuremath{\Upsilon}\xspace}                 
 
 \def\PB      {\ensuremath{\mathrm{B}}\xspace}                 
                  
 \def\PD      {\ensuremath{\mathrm{D}}\xspace}

 \def\PJ      {\ensuremath{\mathrm{J}}\xspace}                 
 \def\PK      {\ensuremath{\mathrm{K}}\xspace}

 \def\Pb      {\ensuremath{\mathrm{b}}\xspace}                 
 \def\Pc      {\ensuremath{\mathrm{c}}\xspace}

 \def\Pi      {\ensuremath{\mathrm{i}}\xspace}

 \def\Ps      {\ensuremath{\mathrm{s}}\xspace}

}
{

 \def\Ppi         {\ensuremath{\pi}\xspace}

 \def\Ppsi        {\ensuremath{\psi}\xspace}                 
                  
 \mathchardef\PDelta="7101
 \mathchardef\PXi="7104
 \mathchardef\PLambda="7103
 \mathchardef\PSigma="7106
 \mathchardef\POmega="710A
 \mathchardef\PUpsilon="7107
                  
 \def\PB      {\ensuremath{B}\xspace}                 
                  
 \def\PD      {\ensuremath{D}\xspace}

 \def\PJ      {\ensuremath{J}\xspace}                 
 \def\PK      {\ensuremath{K}\xspace}

 \def\Pb      {\ensuremath{b}\xspace}                 
 \def\Pc      {\ensuremath{c}\xspace}

 \def\Pi      {\ensuremath{i}\xspace}

 \def\Ps      {\ensuremath{s}\xspace}

}

\makeatletter
\ifcase \@ptsize \relax
  \newcommand{\miniscule}{\@setfontsize\miniscule{4}{5}}
\or
  \newcommand{\miniscule}{\@setfontsize\miniscule{5}{6}}
\or
  \newcommand{\miniscule}{\@setfontsize\miniscule{5}{6}}
\fi
\makeatother

\DeclareRobustCommand{\optbar}[1]{\shortstack{{\miniscule (\rule[.5ex]{1.1em}{.18mm})}
  \\ [-.7ex] $#1$}}
\newcommand{\orbar}[1]{\kern 0.18em\optbar{\kern -0.05em #1}{}\xspace}
 
 \newcommand{\orm}[1]{\shortstack{{\tiny $(-)$}
  \\ [-.7ex] $#1$}}
  
\def\squark    {{\ensuremath{\Ps}}\xspace}

\def\cquark    {{\ensuremath{\Pc}}\xspace}

\def\bquark    {{\ensuremath{\Pb}}\xspace}
\def\bquarkbar {{\ensuremath{\overline \bquark}}\xspace}
\def\bbbar     {{\ensuremath{\bquark\bquarkbar}}\xspace}

\def\pion   {{\ensuremath{\Ppi}}\xspace}

\def\pip    {{\ensuremath{\pion^+}}\xspace}
\def\pim    {{\ensuremath{\pion^-}}\xspace}

\def\kaon    {{\ensuremath{\PK}}\xspace}
  \def\Kbar    {{\kern 0.2em\overline{\kern -0.2em \PK}{}}\xspace}

\def\KorKbar    {\kern 0.18em\optbar{\kern -0.18em K}{}\xspace}

\def\Kp      {{\ensuremath{\kaon^+}}\xspace}
\def\Km      {{\ensuremath{\kaon^-}}\xspace}

\def\Kstarz  {{\ensuremath{\kaon^{*0}}}\xspace}

  \def\Dbar    {{\kern 0.2em\overline{\kern -0.2em \PD}{}}\xspace}
\def\D       {{\ensuremath{\PD}}\xspace}

\def\DorDbar    {\kern 0.18em\optbar{\kern -0.18em D}{}\xspace}
\def\Dz      {{\ensuremath{\D^0}}\xspace}
\def\Dzb     {{\ensuremath{\Dbar{}^0}}\xspace}
\def\Dp      {{\ensuremath{\D^+}}\xspace}
\def\Dm      {{\ensuremath{\D^-}}\xspace}

\def\Dmp     {{\ensuremath{\D^\mp}}\xspace}

\def\Dsp     {{\ensuremath{\D^+_\squark}}\xspace}
\def\Dsm     {{\ensuremath{\D^-_\squark}}\xspace}

\def\Dsmp    {{\ensuremath{\D^{\mp}_\squark}}\xspace}

\def\B       {{\ensuremath{\PB}}\xspace}
\def\Bbar    {{\ensuremath{\kern 0.18em\overline{\kern -0.18em \PB}{}}}\xspace}

\def\BorBbar    {\kern 0.18em\optbar{\kern -0.18em B}{}\xspace}
\def\Bz      {{\ensuremath{\B^0}}\xspace}
\def\Bzb     {{\ensuremath{\Bbar{}^0}}\xspace}
\def\Bu      {{\ensuremath{\B^+}}\xspace}
\def\Bub     {{\ensuremath{\B^-}}\xspace}

\def\Bm      {{\ensuremath{\Bub}}\xspace}

\def\Bmp     {{\ensuremath{\B^\mp}}\xspace}
\def\Bd      {{\ensuremath{\B^0}}\xspace}
\def\Bs      {{\ensuremath{\B^0_\squark}}\xspace}

\def\Bdb     {{\ensuremath{\Bbar{}^0}}\xspace}

\def\jpsi     {{\ensuremath{{\PJ\mskip -3mu/\mskip -2mu\Ppsi\mskip 2mu}}}\xspace}

  \def\Y#1S{\ensuremath{\PUpsilon{(#1S)}}\xspace}

\def\Lz          {{\ensuremath{\PLambda}}\xspace}
\def\Lbar        {{\ensuremath{\kern 0.1em\overline{\kern -0.1em\PLambda}}}\xspace}
\def\LorLbar    {\kern 0.18em\optbar{\kern -0.18em \PLambda}{}\xspace}

\def\Lb      {{\ensuremath{\Lz^0_\bquark}}\xspace}

\def\Lc      {{\ensuremath{\Lz^+_\cquark}}\xspace}

\def\to                 {\ensuremath{\rightarrow}\xspace}

\def\CP                {{\ensuremath{C\!P}}\xspace}

\newcommand{\dm}{{\ensuremath{\Delta m}}\xspace}

\newcommand{\DG}{{\ensuremath{\Delta\Gamma}}\xspace}

\newcommand{\dSS}{\ensuremath{d_{\mathrm{SS}}\xspace}}
\newcommand{\dOS}{\ensuremath{d_{\mathrm{OS}}\xspace}}
\newcommand{\etaSS}{\ensuremath{\eta_{\mathrm{SS}}\xspace}}
\newcommand{\etaOS}{\ensuremath{\eta_{\mathrm{OS}}\xspace}}
\newcommand{\vect}[1]{\ensuremath{\vec{#1}}\xspace}
\newcommand\given[1][]{\:#1\vert\:}

\def\AT#1     {\ensuremath{A_{\mathrm{T}}^{#1}}\xspace}           

\def\C#1      {\ensuremath{\mathcal{C}_{#1}}\xspace}                       
\def\Cp#1     {\ensuremath{\mathcal{C}_{#1}^{'}}\xspace}                    
\def\Ceff#1   {\ensuremath{\mathcal{C}_{#1}^{\mathrm{(eff)}}}\xspace}        
\def\Cpeff#1  {\ensuremath{\mathcal{C}_{#1}^{'\mathrm{(eff)}}}\xspace}       
\def\Ope#1    {\ensuremath{\mathcal{O}_{#1}}\xspace}                       
\def\Opep#1   {\ensuremath{\mathcal{O}_{#1}^{'}}\xspace}                    

\newcommand{\tev}{\ifthenelse{\boolean{inbibliography}}{\ensuremath{~T\kern -0.05em eV}\xspace}{\ensuremath{\mathrm{\,Te\kern -0.1em V}}}\xspace}
\newcommand{\gev}{\ensuremath{\mathrm{\,Ge\kern -0.1em V}}\xspace}
\newcommand{\mev}{\ensuremath{\mathrm{\,Me\kern -0.1em V}}\xspace}
\newcommand{\kev}{\ensuremath{\mathrm{\,ke\kern -0.1em V}}\xspace}
\newcommand{\ev}{\ensuremath{\mathrm{\,e\kern -0.1em V}}\xspace}
\newcommand{\gevc}{\ensuremath{{\mathrm{\,Ge\kern -0.1em V\!/}c}}\xspace}
\newcommand{\mevc}{\ensuremath{{\mathrm{\,Me\kern -0.1em V\!/}c}}\xspace}
\newcommand{\gevcc}{\ensuremath{{\mathrm{\,Ge\kern -0.1em V\!/}c^2}}\xspace}
\newcommand{\gevgevcccc}{\ensuremath{{\mathrm{\,Ge\kern -0.1em V^2\!/}c^4}}\xspace}
\newcommand{\mevcc}{\ensuremath{{\mathrm{\,Me\kern -0.1em V\!/}c^2}}\xspace}

\def\mum  {\ensuremath{{\,\upmu\mathrm{m}}}\xspace}

\def\invfb   {\ensuremath{\mbox{\,fb}^{-1}}\xspace}

\def\fs   {\ensuremath{\mathrm{ \,fs}}\xspace}

\newcommand{\stat}{\ensuremath{\mathrm{\,(stat)}}\xspace}
\newcommand{\syst}{\ensuremath{\mathrm{\,(syst)}}\xspace}

\newcommand{\chisq}{\ensuremath{\chi^2}\xspace}

\newcommand{\chisqip}{\ensuremath{\chi^2_{\text{IP}}}\xspace}

\def\gsim{{~\raise.15em\hbox{$>$}\kern-.85em
          \lower.35em\hbox{$\sim$}~}\xspace}
\def\lsim{{~\raise.15em\hbox{$<$}\kern-.85em
          \lower.35em\hbox{$\sim$}~}\xspace}

\def\ptot       {\mbox{$p$}\xspace}
\def\pt         {\mbox{$p_{\mathrm{ T}}$}\xspace}

\def\degrees{\ensuremath{^{\circ}}\xspace}

\def\evtgen     {\mbox{\textsc{EvtGen}}\xspace}

\def\geant      {\mbox{\textsc{Geant4}}\xspace}

\def\photos     {\mbox{\textsc{Photos}}\xspace}

\def\pythia     {\mbox{\textsc{Pythia}}\xspace}

\def\tell1  {TELL1\xspace}
\def\ukl1   {UKL1\xspace}

\newcommand{\ie}{\mbox{\itshape i.e.}\xspace}

\usepackage{cite} 
\usepackage{mciteplus}

\usepackage[utf8]{inputenc}

\usepackage{longtable} 
\usepackage[colorinlistoftodos]{todonotes}

\newcommand{\BdDpi}{\mbox{\ensuremath{\Bd\to\D^\mp\pi^\pm}}\xspace}
\newcommand{\BdDK}{\mbox{\ensuremath{\Bd\to\D^\mp K^\pm}}\xspace}
\newcommand{\Sf}{\mbox{\ensuremath{S_f}}\xspace}
\newcommand{\Sfb}{\mbox{\ensuremath{S_{\bar{f}}}}\xspace}

\newcommand{\Sfval}{\mbox{\ensuremath{0.058}}\xspace}
\newcommand{\Sfstat}{\mbox{\ensuremath{0.020}}\xspace}
\newcommand{\Sfsyst}{\mbox{\ensuremath{0.011}}\xspace}

\newcommand{\Sfbval}{\mbox{\ensuremath{0.038}}\xspace}
\newcommand{\Sfbstat}{\mbox{\ensuremath{0.020}}\xspace}
\newcommand{\Sfbsyst}{\mbox{\ensuremath{0.007}}\xspace}

\newcommand{\gammaLO}{\mbox{\ensuremath{[5,86]}}\xspace}
\newcommand{\gammaHi}{\mbox{\ensuremath{[185,266]}}\xspace}
\newcommand{\gammaCL}{\mbox{\ensuremath{\gammaLO\degrees \cup \gammaHi\degrees}}\xspace}

\newcommand{\deltaLO}{\mbox{\ensuremath{[-41,41]}}\xspace}
\newcommand{\deltaHi}{\mbox{\ensuremath{[140,220]}}\xspace}
\newcommand{\deltaCL}{\mbox{\ensuremath{\deltaLO\degrees \cup \deltaHi \degrees}}\xspace}

\newcommand{\magSinTwoBplusGCL}{\mbox{\ensuremath{[0.77,1.0]}}\xspace}

\usepackage{booktabs,siunitx}
\usepackage{dcolumn}

\usepackage{bm}

\begin{document}

\renewcommand{\thefootnote}{\fnsymbol{footnote}}
\setcounter{footnote}{1}


\begin{titlepage}
\pagenumbering{roman}

\vspace*{-1.5cm}
\centerline{\large EUROPEAN ORGANIZATION FOR NUCLEAR RESEARCH (CERN)}
\vspace*{1.5cm}
\noindent
\begin{tabular*}{\linewidth}{lc@{\extracolsep{\fill}}r@{\extracolsep{0pt}}}
\ifthenelse{\boolean{pdflatex}}
{\vspace*{-1.5cm}\mbox{\!\!\!\includegraphics[width=.14\textwidth]{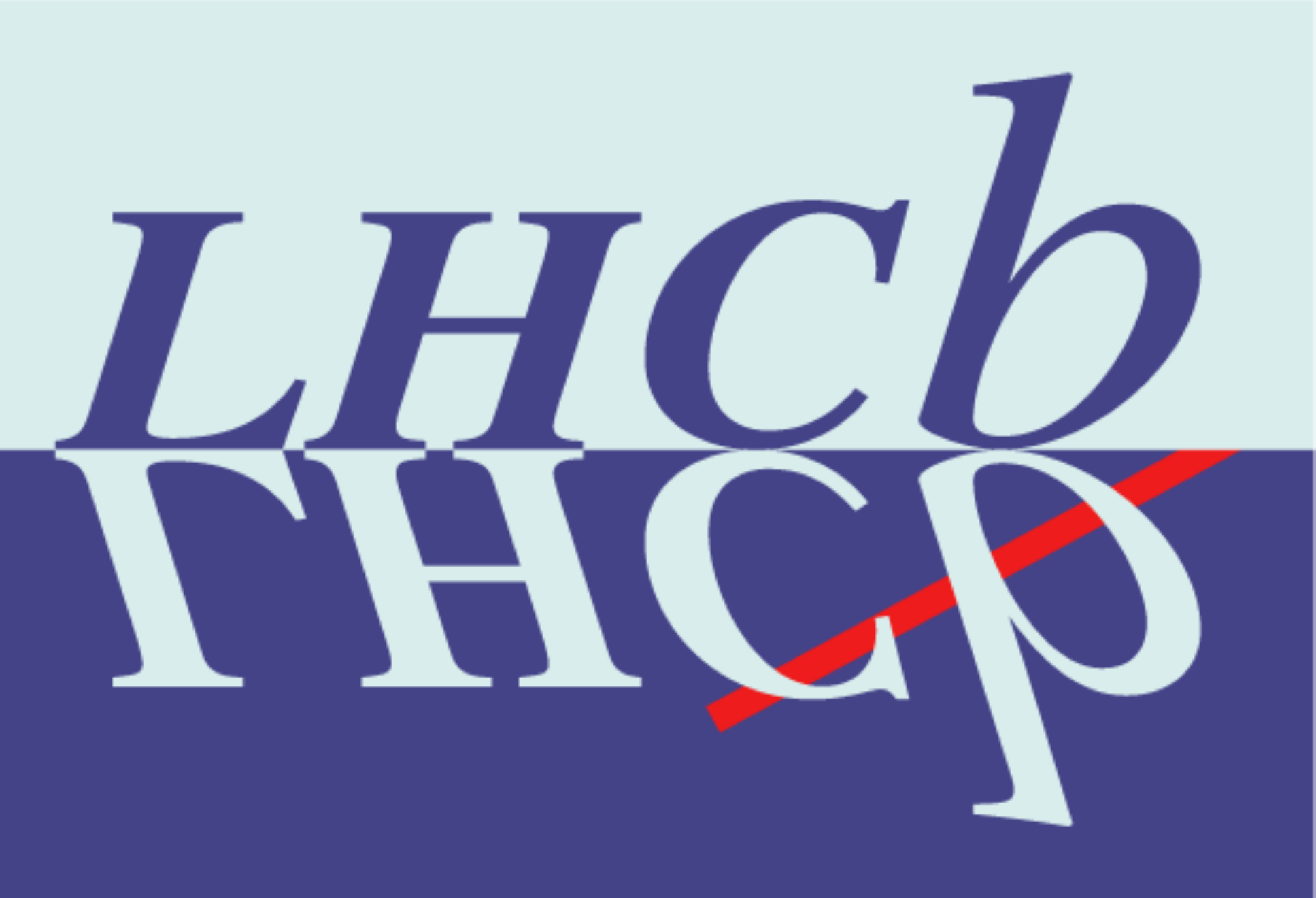}} & &}%
{\vspace*{-1.2cm}\mbox{\!\!\!\includegraphics[width=.12\textwidth]{lhcb-logo.eps}} & &}%
\\
 & & CERN-EP-2018-084 \\  
 & & LHCb-PAPER-2018-009 \\  
 & & \today \\ 
 & & \\
\end{tabular*}

\vspace*{4.0cm}

{\normalfont\bfseries\boldmath\huge
\begin{center}
  \papertitle
\end{center}
}

\vspace*{2.0cm}

\begin{center}
\paperauthors\footnote{Authors are listed at the end of this paper.}
\end{center}

\vspace{\fill}

\begin{abstract}
  \noindent
  A measurement of the \CP asymmetries \Sf and \Sfb in \BdDpi decays is reported.
  The decays are reconstructed in a dataset collected with the \lhcb experiment
  in proton-proton collisions at centre-of-mass energies of 7 and 8\tev  and
  corresponding to an integrated luminosity of 3.0\invfb. The \CP asymmetries are measured
  to be $\Sf = \Sfval \pm \Sfstat\stat \pm \Sfsyst \syst$
  and $\Sfb = \Sfbval \pm \Sfbstat\stat \pm \Sfbsyst \syst$.
  These results are in agreement with, and more precise than, previous determinations.
  They are used to constrain angles of the unitarity triangle, $|\sin\left(2\beta+\gamma\right)|$ and
  $\gamma$, to intervals that are consistent with the current world-average values.
 \end{abstract}

\vspace*{2.0cm}

\begin{center}
  Published in JHEP 06 (2018) 084
\end{center}

\vspace{\fill}

{\footnotesize
\centerline{\copyright~\papercopyright. \href{\paperlicenceurl}{\paperlicence} licence.}}
\vspace*{2mm}

\end{titlepage}


\newpage
\setcounter{page}{2}
\mbox{~}
%
%
%
%

\cleardoublepage


\renewcommand{\thefootnote}{\arabic{footnote}}
\setcounter{footnote}{0}



\pagestyle{plain} 
\setcounter{page}{1}
\pagenumbering{arabic}


%


\section{Introduction}
\label{sec:susedusedA set ofrequiremenntroduction}

In the Standard Model, the decays $\Bd \to\Dm\pip$ and $\Bd \to \Dp\pim$
proceed through the $\bar{b} \to \bar{c} u \bar{d}$ and $\bar{b} \to \bar{u} c \bar{d}$
quark transitions, respectively.\footnote{Inclusion of charge conjugate modes is implied
unless explicitly stated.} The relative weak phase between these two decay amplitudes is
$\gamma\equiv\arg(-V_{ud}V_{ub}^{*}/V_{cd}V_{cb}^{*})$. The \Bd meson
can undergo a flavour oscillation before the decay. The amplitude of the direct decay and that
of a decay preceded by an oscillation have a total relative phase difference of $2\beta + \gamma$,
where $\beta \equiv \arg(-V_{cd}V_{cb}^{*}/V_{td}V_{tb}^{*})$. The phases $\beta$ and $\gamma$
are angles of the unitary triangle. Measurements of \CP violation in \BdDpi decays
provide information on these angles.

Decay-time-dependent \CP asymmetries in \BdDpi decays can be measured by analysing the decay
rates as a function of the decay time of \Bd mesons of known initial
flavour~\cite{Dunietz:1987bvsss,Aleksan:1991nh,Fleischer:2003yb}. The ratio of the decay amplitudes,
\mbox{$r_{D\pi} = |A(\Bd \to \Dp\pim)/A(\Bd \to \Dm\pip)|$}, is around $2\%$,
and limits the size of the \CP asymmetries. Given its small value, this ratio needs to be determined from independent
measurements, for example using the branching ratio of $\Bd \to \Dsp \pi^-$ decays
under the assumption of SU(3) flavour symmetry~\cite{Aubert:2008zi,Das:2010be}.

The decay rates of initially produced $\Bd$ mesons to the final states $f=\Dm\pip$ and $\bar{f}=\Dp\pim$
as a function of the $\Bd$-meson decay time, $t$, are given by
\begin{equation}
\label{eq:decayrates}
\begin{aligned}
\Gamma_{\Bd \to f} (t) &\propto  e^{-\Gamma t} \left[ 1 + C_f \cos( \Delta m \, t ) - \Sf \sin( \Delta m \, t ) \right]\,, \\
\Gamma_{\Bd \to \bar{f}} (t) &\propto e^{-\Gamma t} \left[ 1 + C_{\bar{f}} \cos( \Delta m \, t ) - S_{\bar{f}} \sin(\Delta m  \,t ) \right]\,,
\end{aligned}
\end{equation}
where $\Gamma$ is the average \Bd decay width and $\Delta m$ is the \Bd--\Bdb oscillation
frequency. 
For an initially produced \Bdb meson, the same equations hold except for a change of sign of
the coefficients in front of the sine and cosine functions.
No \CP violation in the decay is assumed, \ie only tree-level processes
contribute to the decay amplitudes.
It is also assumed that $|q/p|=1$, where $q$ and $p$ are the complex coefficients defining
the heavy and light mass eigenstates of the \Bd system, and $\Delta\Gamma=0$, where $\Delta\Gamma$
is the decay-width difference between the two mass eigenstates. These assumptions follow from the
known values of these quantities~\cite{HFLAV16}. Under these assumptions, the coefficients of the
cosine and sine terms of Eq.~\eqref{eq:decayrates} are given by
\begin{align}
C_f  &=  \frac{ 1 - r_{D\pi}^2 }{ 1 + r_{D\pi}^2 } = - C_{\bar{f}} \,,  \\
	\Sf  &= -\frac{ 2 r_{D\pi} \sin\left[\delta - (2\beta + \gamma)\right]}{ 1 + r_{D\pi}^2 }    \,, \\
	\Sfb &= \frac{ 2 r_{D\pi} \sin\left[\delta + (2\beta + \gamma)\right]}{ 1 + r_{D\pi}^2 }      \,,
\end{align}
where $\delta$ is the \CP-conserving phase difference between the
$\bar{b} \to \bar{c} u \bar{d}$ and $\bar{b} \to \bar{u} c \bar{d}$ decay amplitudes.
Due to the small value of $r_{D\pi}$, terms of $\mathcal{O}(r_{D\pi}^2)$ are neglected in this analysis, fixing $C_f=-C_{\bar{f}}=1$.

A measurement of the \CP asymmetries \Sf and \Sfb can be interpreted in terms of $2\beta+\gamma$ by using the value
of $r_{D\pi}$ as input. Additionally, using the known value of $\beta$~\cite{HFLAV16},
the angle $\gamma$ can be evaluated.  The determination of $\gamma$ from tree-level decays is
important because processes beyond the Standard Model are not expected to contribute. Constraints from the
analysis of \BdDpi decays can be combined with other measurements to improve the ultimate
sensitivity to this angle~\cite{LHCb-PAPER-2016-032}.

Measurements of \Sf and \Sfb using $\Bd\to D^{(*)\mp}\pi^\pm$ and  $\Bd\to D^{\mp}\rho^\pm$ decays have been reported by the BaBar~\cite{Aubert:2005yf,Aubert:2006tw} and Belle~\cite{PhysRevD.73.092003,Bahinipati:2011yq} collaborations.
This paper presents a measurement of \Sf and \Sfb with \BdDpi decays reconstructed in a dataset
collected with the \lhcb experiment in proton-proton collisions at centre-of-mass energies of \num{7} and \num{8}\tev and corresponding to an integrated luminosity of 3.0\invfb. This is the first measurement of \Sf and \Sfb at a hadron collider.

\section{Detector and simulation}
\label{sec:detector}

The \lhcb detector~\cite{Alves:2008zz,LHCb-DP-2014-002} is a single-arm forward
spectrometer covering the \mbox{pseudorapidity} range 2--5,
designed for the study of particles containing \bquark or \cquark
quarks. The detector includes a high-precision tracking system
consisting of a silicon-strip vertex detector surrounding the $pp$
interaction region~\cite{LHCb-DP-2014-001}, a large-area silicon-strip
detector located upstream of a dipole magnet with a bending power of about
$4{\mathrm{\,Tm}}$, and three stations of silicon-strip detectors and straw
drift tubes~\cite{LHCb-DP-2013-003} placed downstream of the magnet.
The tracking system provides a measurement of the momentum, \ptot, of charged
particles with a relative uncertainty that varies from $0.5\%$ at low momentum to $1.0\%$
at 200\gevc. The minimum distance of a track to a primary vertex (PV), the impact
parameter (IP), is measured with a resolution of $(15+29/\pt)\mum$, where \pt is the
component of the momentum transverse to the beam, in\,\gevc. Different types of charged
hadrons are distinguished using information from two ring-imaging Cherenkov detectors.
Photons, electrons and hadrons are identified by a calorimeter system consisting
of scintillating-pad and preshower detectors, an electromagnetic calorimeter and
a hadronic calorimeter. Muons are identified by a system composed of alternating
layers of iron and multiwire proportional chambers.

In the simulation, $pp$ collisions are generated using
\pythia~\cite{Sjostrand:2006za,*Sjostrand:2007gs} with a specific \lhcb
configuration~\cite{LHCb-PROC-2010-056}. Decays of hadronic particles are
described by \evtgen~\cite{Lange:2001uf}, in which final-state radiation is
generated using \photos~\cite{Golonka:2005pn}. The interaction of the generated
particles with the detector, and its response, are implemented using the \geant
toolkit~\cite{Allison:2006ve, *Agostinelli:2002hh} as described in
Ref.~\cite{LHCb-PROC-2011-006}.


\section{Candidate selection}
\label{sec:selection}

The online event selection is performed by a trigger,
which consists of a hardware stage, using information from the calorimeter and muon
systems, followed by a software stage, which applies a full event
reconstruction. Events containing a muon with high \pt or a
 hadron, photon or electron with high transverse energy in the calorimeters
 are considered  at the hardware trigger stage.
Events selected by the trigger using hadrons from the signal decay
represent $70\%$ of the sample used in this analysis, the rest being collected
using trigger criteria satisfied by other properties of the event.

The software trigger requires a two-, three-, or four-track secondary vertex with a
significant displacement from the primary $pp$ interaction vertices. At least
one charged particle must have $\pt > 1.7\gevc$ and be inconsistent with
originating from a PV. A multivariate algorithm is used for the identification
of secondary vertices consistent with the decay of a $\bquark$ hadron~\cite{BBDT}.

The selection of $\BdDpi$ candidates is performed by reconstructing ${\Dm\to\Kp\pim\pim}$
candidates from charged particle tracks with high momentum and transverse momentum, and
originating from a common displaced vertex. Particle identification (PID) information is
used to select kaon and pion candidates, and the $\Kp\pim\pim$ invariant mass is required
to be within $35\mevcc$ of the known value of the $\Dm$ mass~\cite{PDG2017}. These candidates
are combined with a fourth charged particle, referred to as the \emph{companion}, to form
the $\Bd$ vertex, which must be displaced from any PV. The PV with respect to which the $\Bz$
candidate has the smallest $\chisqip$ is considered as the production vertex. The $\chisqip$ is
defined as the difference in the vertex-fit \chisq of a given PV reconstructed with and
without the \Bz candidate. No PID requirement is applied to the companion track at this stage.

The $\BdDpi$ candidates are required to match the secondary vertices found in
the software trigger, to have a proper decay time larger than \SI{0.2}{ps}, and
to have a momentum vector aligned with the vector formed by joining the PV
and the \Bz decay vertex. The decay time is determined from a kinematic fit in which the $\Bd$ candidate is constrained to
originate from the PV to improve the decay-time resolution, while the $\Bd$-candidate mass is
computed assigning the known value~\cite{PDG2017} to the mass of the $\Dm$ candidate to
improve the mass resolution~\cite{Hulsbergen:2005pu}. A combination of PID information and
mass-range vetoes is used to suppress to a negligible level cross-feed backgrounds such as
$\Lb\to\Lc (\to p \Km \pip) \pim$ and $\Bs\to\Dsm(\to\Km\Kp\pim)\pip$, due to the misidentification
of protons and kaons as pions.

A boosted decision tree~(BDT)~\cite{Breiman,AdaBoost} is used to increase the signal purity by suppressing background from random combinations of particles.
Candidates reconstructed from simulated $\BdDpi$ decays are used as signal in the training of the BDT, and data candidates with an invariant mass larger than $5.5\gevcc$ are used as background.
A set of \num{16} variables are combined into a single response, which is used to categorise the \Bz candidates.
The most relevant variables entering the BDT are the quality of the fit of the \Bz vertex and that of the kinematic fit to calculate the \Bz decay time, the transverse momentum of the $D^-$ candidate, and the quality of the fit of the companion-particle track.
The requirement placed on the BDT response is chosen to maximise the expected sensitivity to $\Sf$ and $\Sfb$ as derived from a
set of simulated samples of signal  plus background that are passed through the entire analysis.
The data sample is further required to consist of $\Bz$ candidates whose initial flavour has been determined by means of the flavour tagging algorithms described in Sec.~\ref{sec:tagging}.


\section{Sample composition}
\label{sec:sample}
\begin{figure}[tb]
	\begin{center}
		\includegraphics[width=0.49\linewidth]{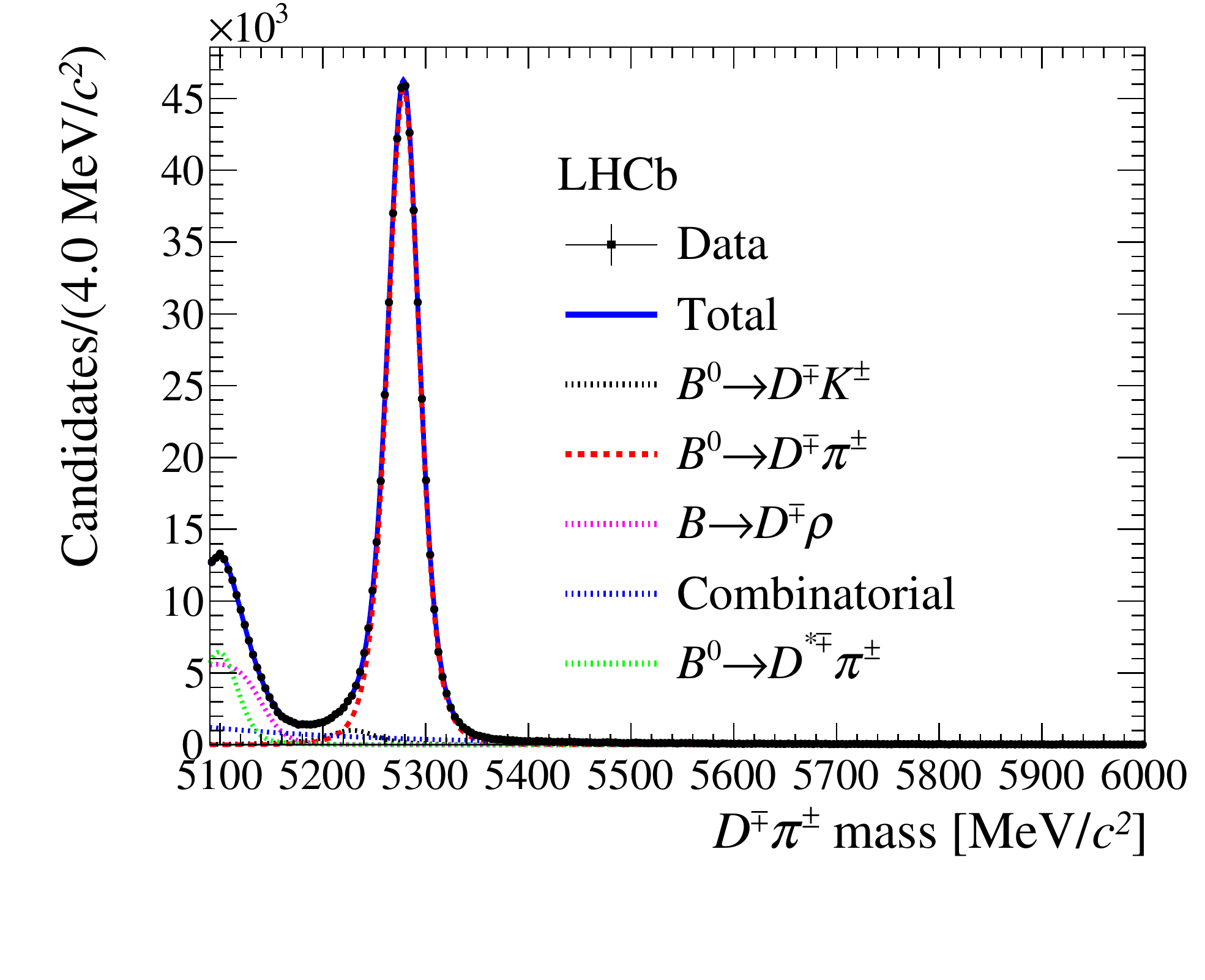}
		\includegraphics[width=0.49\linewidth]{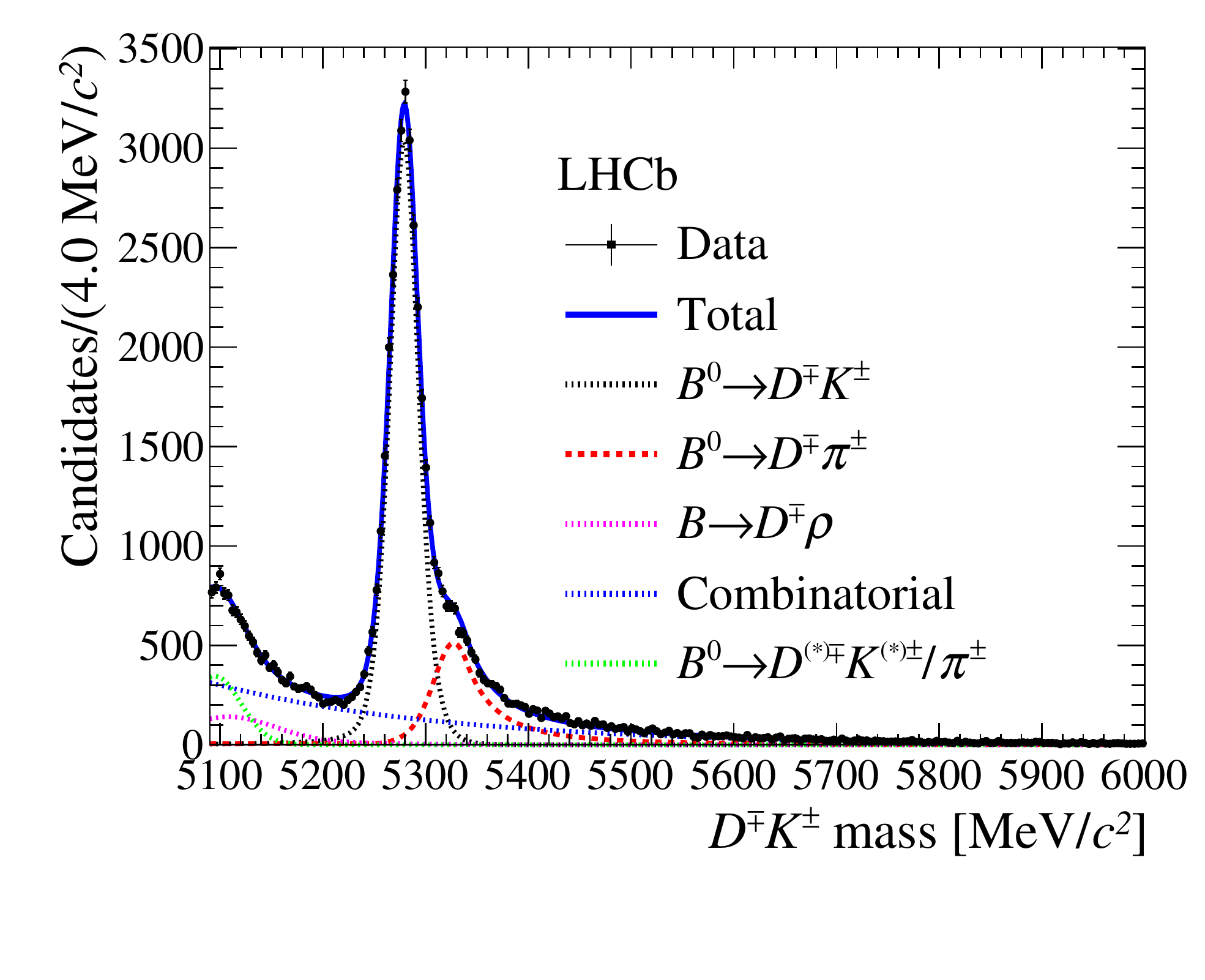}
	\end{center}
	\vspace{-8 mm}
	\caption{Invariant mass distributions of the (left) pion-like and (right) kaon-like samples
	with fit projections overlaid. The simultaneous fit of the two distributions is
	described in the text and yields a $\chi^2$ per degree of freedom of 1.18.
        The $B\to\Dmp\rho$ component includes both $\Bz\to\Dmp\rho^\pm$ and $\Bmp\to\Dmp\rho^0$ decays.}
	\label{fig:mass_fit}
\end{figure}

The data sample after the selection is split into two disjoint subsets according
to the PID information of the companion particle: a sample referred to as
\emph{pion-like} consisting mostly of genuine $\BdDpi$ decays,
and a sample referred to as \emph{kaon-like} consisting mostly of genuine $\BdDK$ decays.
The binned $\Bd$-mass distributions of these
two samples are fitted simultaneously in order to determine the sample compositions. The  mass
distributions span the range 5090--6000\mevcc and are shown in Fig.~\ref{fig:mass_fit} with fit
projections overlaid.

The mass distribution of $\Bd$ candidates in the pion-like sample features a peak at the
known $\Bd$ mass with a width of about \SI{20}{\mevcc}, corresponding to $\BdDpi$
signal decays, and is modelled with the sum of a double-sided Hypatia
function~\cite{Santos:2013ky} and a Johnson SU function~\cite{JohnsonSU}. The
combinatorial background is modelled using the sum of two exponential functions.
At values lower than \SI{5.2}{\gevcc}, broad structures
corresponding to partially reconstructed decays, such as $\Bd\to\Dm\rho^+(\to \pi^+ \pi^0)$, $\Bm\to\Dm\rho^0(\to \pip \pim)$
and $\Bd\to D^{*-} (\to\Dm\pi^0)\pip$ where the additional pion is not reconstructed,
are present; the shapes of these backgrounds are determined from simulation.
Cross-feed $\BdDK$ decays, due to kaon-to-pion misidentification, contaminating the left tail of the signal peak, are described
with a double-sided Hypatia function with parameters determined from simulated decays.

The $\Bd$-mass distribution of the kaon-like sample contains analogous
components: the $\BdDK$ signal peak is modelled with a single-sided Hypatia function;
the combinatorial background with an exponential function;
partially reconstructed \mbox{$\Bd \to \Dm \rho^+(\to \pi^+ \pi^0)$},
$\Bd \to D^{*-}  (\to \Dm \pi^0) \pip$, $\Bd \to D^{*-}  (\to \Dm \pi^0) K^+$ and
$\Bd \to \Dm K^{*+} (\to \pi^0 K^+)$ decays, where the charged pion is misidentified as
a kaon and the neutral pion is not reconstructed, are modelled using simulation.
Cross-feed $\BdDpi$ decays from pion-to-kaon misidentification in the kaon-like sample peaks to the right of the
$\BdDK$ signal region, with a long tail towards the high-mass region; the shape of this
distribution, a double-sided Hypatia function, is taken from simulation.

The yields of all components are floating parameters of the fit. The yield of
the $\BdDK$ cross-feed decays in the pion-like sample is constrained to that of
the $\BdDK$ signal decays in the kaon-like sample using
the kaon-to-pion misidentification probability and the kaon identification efficiency of the PID requirement
on the companion particle. In a similar manner, the yield
of the $\BdDpi$ cross-feed decays in the kaon-like sample is constrained to that of
$\BdDpi$ signal decays in the pion-like sample scaled by the
pion-to-kaon misidentification probability and the pion identification efficiency.
The misidentification probabilities and the identification efficiencies are determined
from a large sample of $D^{*+}\to D^0(\to K^-\pip)\pip$
decays in which the charged tracks are weighted in momentum and pseudorapidity to
match those of the companion particle in $\BdDpi$ decays~\cite{LHCb-PUB-2016-021}.

An unbinned maximum-likelihood fit to the $\Bd$-mass distribution of the pion-like sample
is performed to determine \emph{sWeights} \cite{Pivk:2004ty},
which are used to statistically subtract
the background in the decay-time analysis of Sec.~\ref{sec:fit}.
This unbinned fit contains the same components as the binned fit,
but applied in a smaller mass window, 5220--5600\mevcc, to suppress the background contamination.
All backgrounds entering this mass region are combined to
form a single shape according to the fractions found in the previous fit. The shape parameters
of the signal and background components are also fixed to the values found in the preceding fit.
The \BdDpi signal yield is found to be \num[separate-uncertainty=true]{479000\pm700}
and that of the background  to be \num[separate-uncertainty=true]{34400\pm300}.


\section{Flavour tagging}
\label{sec:tagging}

A combination of tagging algorithms is used to determine the flavour of the \Bz candidates
at production. Each algorithm provides a decision (tag), $d$, which determines the flavour, and an
estimate, $\eta$, of the probability that the decision is incorrect (mistag probability).
The decision takes the value of $d=1$ for a candidate tagged as a \Bz, and $d=-1$ for
a candidate tagged as \Bzb. The mistag probability is defined only between 0 and 0.5,
since $\eta>0.5$ corresponds to an opposite tag with a mistag probability of $(1-\eta)$.

Two classes of flavour tagging algorithms are used: opposite-side, OS, and same-side, SS,
taggers. The OS tagger exploits the dominant production
mechanism of $\bquark$ hadrons, the incoherent production of $\bbbar$ pairs, by
identifying signatures of the $\bquark$ hadron produced together with the signal $\Bd$
meson. The time evolution of the signal $\Bd$ meson is independent from that of the accompanying $\bquark$ hadron. 
The OS tagger uses the charge of the electron or muon from semileptonic
$\bquark$-hadron decays, the charge of the kaon from a $b\to c \to s$ decay chain,
the charge of a reconstructed secondary charm hadron, and the charge
of particles associated with a secondary vertex distinct from the signal decay;
further details are given in Refs.~\cite{LHCb-PAPER-2011-027,LHCb-PAPER-2015-027}.

The SS tagger selects pions and protons related to the hadronisation process of the signal $\Bd$
meson by means of BDT classifiers that determine the tag decision and mistag
probability, as described in Ref.~\cite{LHCb-PAPER-2016-039}. Unlike
Ref.~\cite{LHCb-PAPER-2016-039}, where $\BdDpi$ decays are used assuming $S_{f}=S_{\bar f}=0$,
the BDT classifiers of the SS algorithm
exploited in this analysis are trained on a control sample of flavour-specific
$\Bd\to\jpsi\Kstarz$ decays, whose distributions of $\pt$, pseudorapidity,
azimuthal angle of the $\Bd$ candidate, as well as number of tracks and PVs in the
event, are weighted to match those of the $\BdDpi$ signal decay.

Around $37\%$ of the \Bd candidates are tagged by the OS tagger, $79\%$
by the SS tagger, and $31\%$ by both algorithms. About $15\%$ of the \Bz
candidates are not tagged by either of the algorithms and are discarded. Each tagging
decision is weighted by the estimated mistag probability $\eta$, which dilutes the sensitivity
to the \CP asymmetry. To correct for potential biases in $\eta$, a function \mbox{$\omega\hspace{0.5mm}(\eta)$} is
used to calibrate the mistag probability which provides an unbiased estimate of the mistag fraction
$\omega$ ($\bar{\omega}$), \ie the fraction of incorrectly tagged candidates for a
\Bz (\Bzb) meson, for any value of $\eta$.

Charged particles used for flavour tagging, such as the kaons from the $b\to c \to s$ decay
chain exploited in the OS tagger, can have different interaction cross-sections with
the detector material and therefore different reconstruction efficiencies. This can result
in different tagging efficiencies and mistag probabilities for initial $\Bd$ and $\Bdb$
mesons. Asymmetries in the tagging efficiency are found to be consistent
with zero in simulation and data for both taggers and are therefore neglected in the baseline fit,
but considered as a source of systematic uncertainty. This is not the case for the
asymmetries of the mistag probability, which can bias the determination of the $\CP$ asymmetries
and must be corrected for. Therefore, the calibration functions depend
on the initial flavour of the \Bd candidate: $\omega(\eta)$ for $d=+1$ and  $\overline{\omega}(\eta)$ for $d=-1$.
They are expressed as generalised linear models (GLMs) of the form
\begin{equation}
  \label{eq:glm_def}
  \orbar{\omega}(\eta) = g\big(h(\eta)\big) = g\bigg( g^{-1} (\eta) + \sum\limits_{i=1}^{N} \left(p_i \orm{+} \frac{\Delta p_i}{2}\right) f_i(\eta)\bigg)\,,
\end{equation}
where $p_i$ and $\Delta p_i$ are free parameters, $f_i$ are the {\it{basis functions}}, and $g$ is the {\it{link function}}~\cite{GLM}.

The calibration function of the OS tagger is a GLM using natural splines as the basis
functions~\cite{Nsplines} with five \emph{knots}, $N=5$. For the SS tagger, a GLM using first-order
polynomial basis functions and $N=2$ is used. In both cases a modified logistic function,
$g(x) = \frac{1}{2}(1+e^x)^{-1}$, is used as the link function.
To account for the tagging decision and mistag probability,
the following substitutions occur in Eq.~\eqref{eq:decayrates}:
\begin{equation}
\label{eq:with_tagging}
\begin{split}
S_{f} &\to (\Delta^- - \Delta^+)S_f\,, \\
C_{f} &\to (\Delta^- - \Delta^+)C_f\,.
\end{split}
\end{equation}
Similar equations hold for \Sfb and $C_{\bar{f}}$.
The calibration functions enter the coefficients $\Delta^\pm$ along with
the tagging efficiencies $\varepsilon_\text{OS}$ and $\varepsilon_\text{SS}$ of the OS and SS taggers, according to
\begin{align}
\Delta^\pm & =  \frac{1}{2} \varepsilon_\text{\tiny{OS}} \bigg[1-\varepsilon_\text{\tiny{SS}}+d_\text{\tiny{OS}}\Big(1-\varepsilon_\text{\tiny{SS}}-2\omega(\eta_\text{\tiny{OS}})\big(1 +\varepsilon_\text{\tiny{SS}}\big)\Big)\bigg] \nonumber\\
  & \pm  \frac{1}{2} \varepsilon_\text{\tiny{OS}} \bigg[1-\varepsilon_\text{\tiny{SS}}+d_\text{\tiny{OS}}\Big(1- \varepsilon_\text{\tiny{SS}}-2\overline{\omega}(\eta_\text{\tiny{OS}})\big(1 +\varepsilon_\text{\tiny{SS}}\big)\Big)\bigg]\,,
\end{align}
for candidates tagged by the OS algorithm and not by the SS algorithm (and vice-versa, exchanging the OS and SS indexes), and
\begin{align}
\Delta^\pm &= \frac{1}{4} \varepsilon_\text{\tiny{OS}} \varepsilon_\text{\tiny SS} \left[ 1 +\hspace{-3mm}\sum\limits_{j=\text{\tiny OS},\text{\tiny SS}}\hspace{-2mm} d_j\Big(1-2\omega(\eta_j)\Big) + d_\text{\tiny OS}d_\text{\tiny SS}\Big(1 - 2\omega(\eta_j) + 2\omega(\eta_\text{\tiny OS})\omega(\eta_\text{\tiny SS})\Big) \right] \nonumber \\
& \pm \frac{1}{4} \varepsilon_\text{\tiny{OS}} \varepsilon_\text{\tiny SS} \left[ 1 +\hspace{-3mm}\sum\limits_{j=\text{\tiny OS},\text{\tiny SS}}\hspace{-2mm} d_j\Big(1-2\overline{\omega}(\eta_j)\Big) + d_\text{\tiny OS}d_\text{\tiny SS}\Big(1 - 2\overline{\omega}(\eta_j) + 2\overline{\omega}(\eta_\text{\tiny OS})\overline{\omega}(\eta_\text{\tiny SS})\Big) \right] \,,
\end{align}
for candidates tagged by both algorithms.
The form of the $\Delta^\pm$ coefficients and of the substitutions of Eq.~\eqref{eq:with_tagging} is convenient to also account
for other spurious asymmetries considered in Sect.~\ref{sec:fit}.

\begin{figure}[tb]
\includegraphics[width=0.45\textwidth]{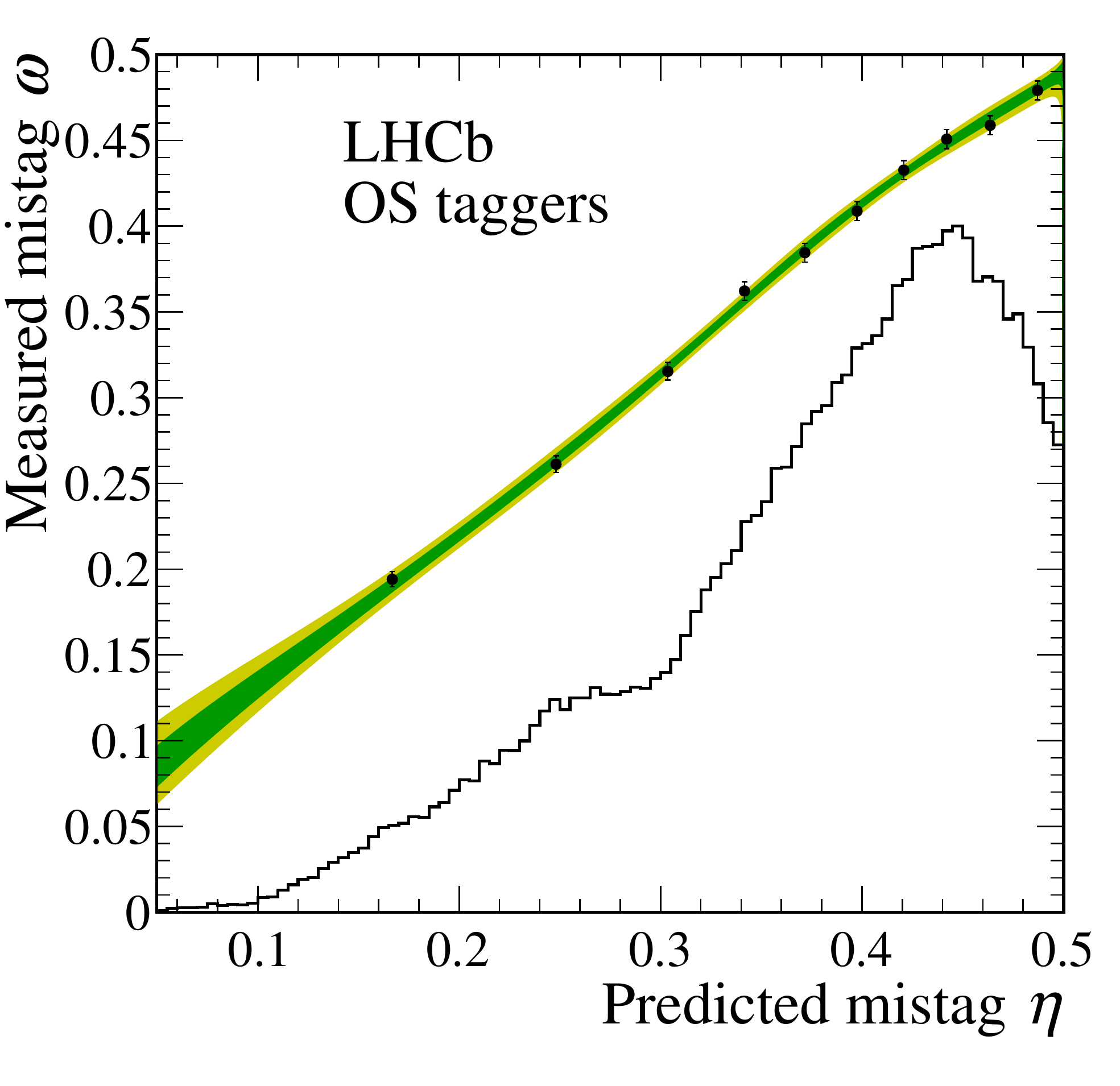}\hfil
\includegraphics[width=0.45\textwidth]{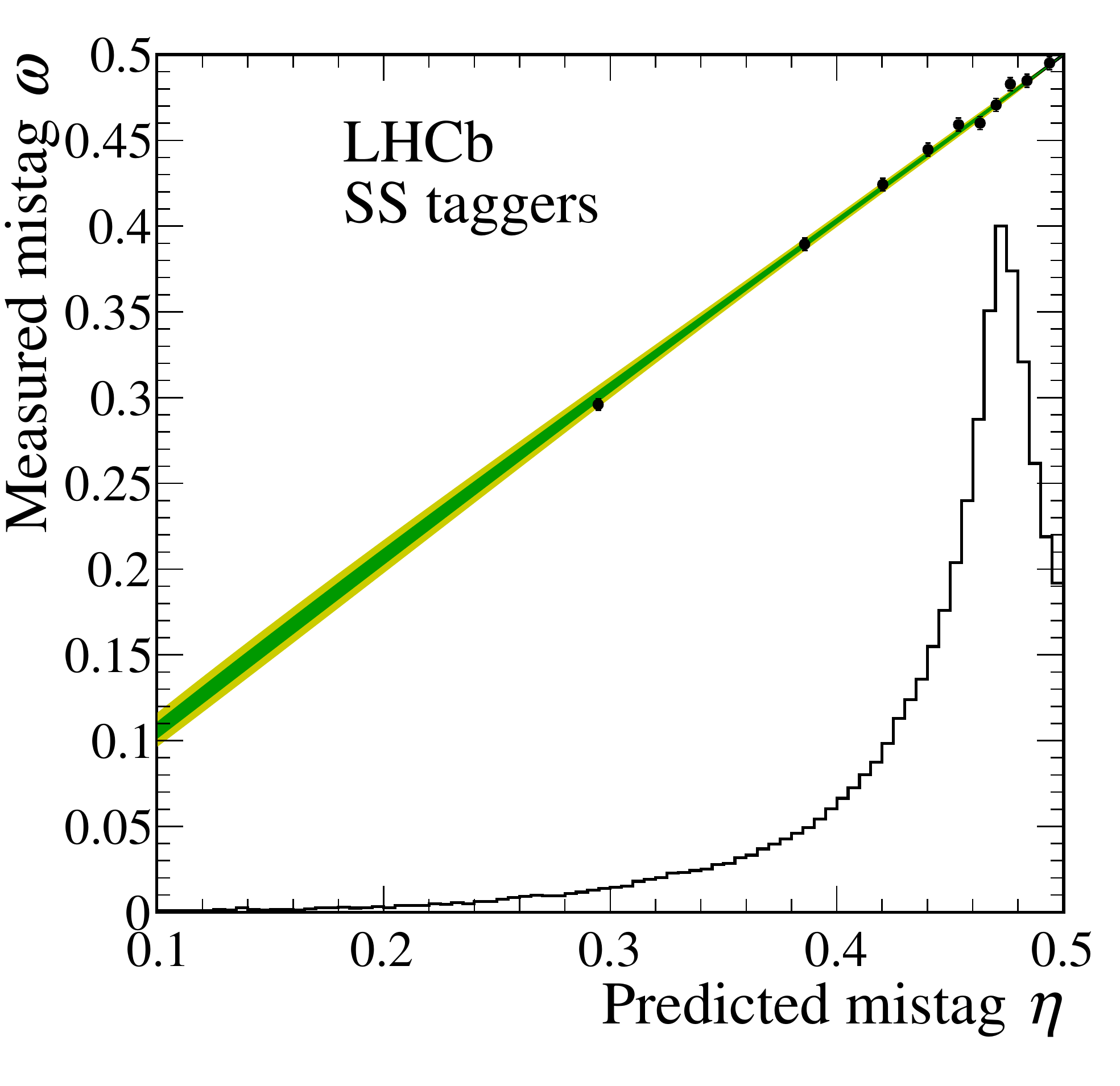}
\caption{Measured mistag fraction $\omega$ versus predicted mistag probability  $\eta$ of the combination of
(left) OS and (right) SS taggers as determined in signal decays with the fit described in
Sect.~\ref{sec:fit}. The black histograms are the distributions of the mistag probabilities
in arbitrary units. The shaded areas correspond to the 68\% and 95\%
confidence-level regions of the calibration functions and do not include systematic uncertainties
on the parameters. The calibration functions and the distributions of mistag probabilities are
shown summing over candidates tagged as either \Bz or \Bzb.}
\label{fig:tagging_calib}
\end{figure}

The seven pairs of calibration parameters $(p_i, \Delta p_i)$ are left free in the fit
from which the \Sf and \Sfb observables are extracted. This is possible
because the $C_f$ and $C_{\bar{f}}$ coefficients are fixed parameters,
so that the cosine terms of the decay rates permit the calibration
parameters to be measured.
This procedure has been validated with pseudoexperiments and possible deviations of $C_f$ and $C_{\bar{f}}$ from unity are taken into account in the systematic uncertainties.
To account for possible mismodelling of the calibration functions, systematic uncertainties are assigned to $S_f$ and $S_{\bar{f}}$.
The calibration functions obtained in the data
are shown in Fig.~\ref{fig:tagging_calib}, where the
measured mistag fraction is presented as a function
of the predicted mistag probability of the tagger.

Considering only candidates retained for the analysis, \ie those with a flavour tag, the statistical uncertainties of $\Sf$ and $\Sfb$ are inversely proportional to
$\sqrt{\langle\mathcal{D}^2\rangle}$.  Here,  $\langle\mathcal{D}^2\rangle$ is the average of the squared dilution of the signal,
calculated as $\frac{1}{\mathcal{N}_{\mathrm{tag}}}\sum_{i=1}^{N_{\mathrm{tag}}}w_i\left[1-2\omega(\eta_{i})\right]^2$,
where $N_{\mathrm{tag}}$ is the number of candidates, $w_i$ is the $sWeight$ of the candidate $i$ determined
in the fit of the sample composition, and $\mathcal{N}_{\mathrm{tag}}=\sum_{i=1}^{N_{\mathrm{tag}}}w_i$.
The total dilution squared of the sample is found to be $(6.554 \pm 0.017)\%$.
Considering also the number of discarded candidates because no tagging decision is determined by either tagger,
$N_{\mathrm{untag}}$ and $\mathcal{N}_{\mathrm{untag}}=\sum_{i=1}^{N_{\mathrm{untag}}}w_i$, the tagging efficiency
$\varepsilon_{\rm tag}\equiv \mathcal{N}_{\mathrm{tag}}/(\mathcal{N}_{\mathrm{tag}}+\mathcal{N}_{\mathrm{untag}})$ is found to be $(85.23 \pm 0.05)\%$.
 Hence, the effective tagging efficiency of the initial sample is $\varepsilon_{\rm tag}\langle\mathcal{D}^2\rangle=(5.59\pm0.01)\%$.
 All quoted uncertainties are statistical only. The effective tagging efficiency is similar to that of
 the measurement of \CP violation in $\Bs \to \Dsmp K^\pm$ decays~\cite{LHCb-PAPER-2017-047}.

\section{Decay-time fit}
\label{sec:fit}

The $\CP$ asymmetries $\Sf$ and $\Sfb$ are determined from a
multidimensional maximum-likelihood fit to the unbinned distributions
of the signal candidates weighted with the \emph{sWeights}. The probability density function
(PDF) describing the signal decay to a final state $F$ equal to $f$  or  $\bar{f}$, at
the reconstructed decay time $t$, and given the tags
$\vect{d} = (\dOS, \dSS)$ and mistag probabilities
$\vect{\eta} = (\etaOS, \etaSS)$, is
\begin{equation}
	P(t, F, \vect{d} \given \vect{\eta}) \propto \epsilon(t) \left(\mathcal{P}(t', F, \vect{d} \given \vect{\eta})
    \otimes \mathcal{R}(t'-t)\right)\,,
\end{equation}
where  $\mathcal{P}(t',F,\vect{d} \given \vect{\eta})$ is the function describing the
distribution of true decay times $t'$, $\mathcal{R}(t'-t)$ is the decay time
resolution, and $\epsilon(t)$ describes the decay-time-dependent efficiency of
reconstructing and selecting the signal decays. The function $\mathcal{P}(t', F, \vect{d} \given \vect{\eta})$
corresponds to one of the decay rates of Eq.~\eqref{eq:decayrates}, according to the final state $F$,
and with the substitutions of Eq.~\eqref{eq:with_tagging} to include the flavour tagging.

A  production asymmetry,
$A_\text{P}$, and a final-state detection asymmetry, $A_\text{D}$, must also be taken into account.
These are defined as
\begin{equation}
A_{\rm{P}} = \frac{\sigma(\Bzb) - \sigma(\Bz)}{\sigma(\Bzb) + \sigma(\Bz)} \,, \qquad
A_{\rm{D}}  = \frac{\varepsilon(f) - \varepsilon(\bar{f})}{\varepsilon(f) + \varepsilon(\bar{f})}\,,
\end{equation}
where $\varepsilon$ is the decay-time-integrated efficiency
in reconstructing and selecting the final state $\bar{f}$ or $f$, and
$\sigma$ is the production cross-section of the given \Bzb or \Bz meson.
The asymmetry $A_{\rm{P}}$ arises from the different production cross-sections of \Bzb and \Bz
mesons in proton-proton collisions and is measured to be at the percent level at LHC energies \cite{LHCb-PAPER-2016-062}.
The detection asymmetry is also measured to be at the percent level and to
be independent of the decay time. Therefore, Eq.~\eqref{eq:with_tagging} is further modified as follows:
\begin{equation}
\label{eq:coeff_complete}
\begin{split}
(\Delta^- - \Delta^+) S_f &\to (\Delta^- - A_{\rm{P}} \Delta^+)(1 + A_{\rm{D}}) S_f\,, \\
(\Delta^- - \Delta^+) C_f &\to  (\Delta^- - A_{\rm{P}} \Delta^+)(1 + A_{\rm{D}}) C_f\,,
\end{split}
\end{equation}
where $C_f$ is fixed to 1.
Similar equations hold for \Sfb and $C_{\bar{f}}$ (fixed to $-1$) with $A_{\rm{D}} \to -A_{\rm{D}}$.

The decay-time resolution is determined from a sample of {\em fake} $\Bz$ candidates
formed from a genuine $\Dm$ meson and a charged track originating from the same PV
and consistent with being a pion of opposite charge. These candidates are subjected
to a selection similar to that of the signal decays except for all decay-time biasing
requirements, which are removed. The decay-time distribution of these candidates
is therefore expected to peak at zero with a Gaussian shape given by the resolution function.
Its width is determined in bins of the uncertainty on the decay time provided by the kinematic fit
of the decay chain. A second-order polynomial is used to describe the measured width as a function of the
decay-time uncertainty. The average resolution of \SI[separate-uncertainty=true]{54.9\pm0.4}{\fs}
is used as the width of the Gaussian resolution function $\mathcal{R}(t'-t)$.
The efficiency function $\epsilon(t)$ is modelled by segments of cubic
b-splines~\cite{splines} with nine free parameters in total.

The free parameters of the fit are the $S_{f}$ and $S_{\bar f}$ coefficients,
the detection and production asymmetries $A_{\rm{D}}$ and $A_{\rm{P}}$,
the seven pairs of parameters $(p_i, \Delta p_i)$ for the calibration functions
of the OS and SS taggers, their efficiencies $\varepsilon_\text{OS}$ and $\varepsilon_\text{SS}$,
and the nine  parameters of $\epsilon(t)$. The average \Bz decay width, $\Gamma$ in Eq.~\eqref{eq:decayrates},
is constrained by means of a Gaussian function whose mean is the world average value and whose width
is the uncertainty~\cite{HFLAV16}. Similarly, the \Bz--\Bzb mixing frequency, \dm, is constrained to the value
measured in Ref.~\cite{LHCB-PAPER-2015-031}.

\begin{figure}[tb]
 	\centering
	\includegraphics[width=0.5\textwidth]{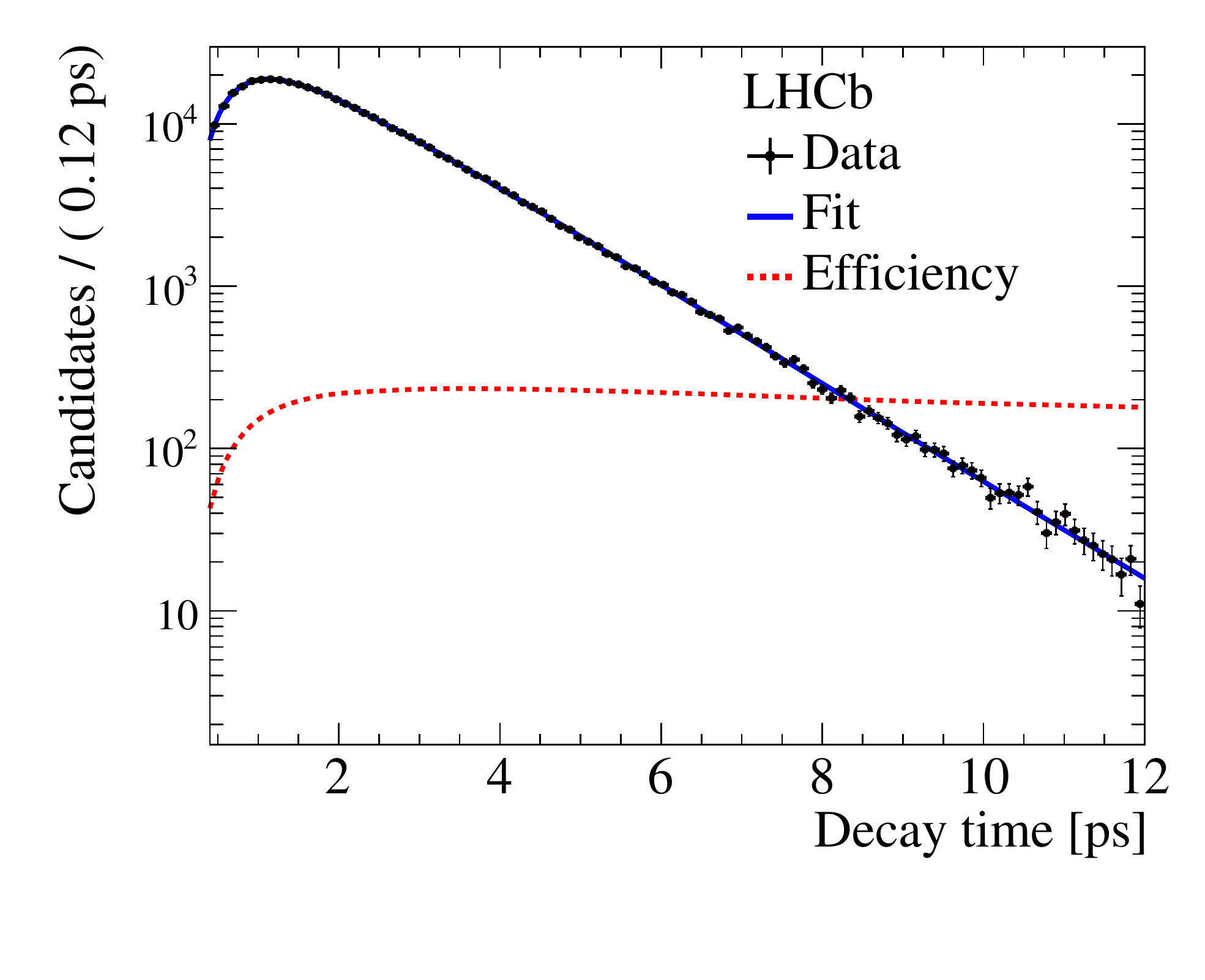}
\caption{
		Background-subtracted decay-time distribution for tagged candidates.
	The solid blue curve is the projection of the signal PDF. The red dotted
	curve indicates the efficiency function $\varepsilon\left(t\right)$ in arbitrary units.}
\label{fig:decaytime}
\end{figure}

The fit determines $\Sf = \Sfval \pm 0.021$ and $\Sfb = \Sfbval \pm 0.021$ where the uncertainties include the contributions from the constraints on the decay width and mixing frequency.
When the fit is repeated by fixing $\dm$ and $\Gamma$ to the
central values used in the constraints, the central values for \Sf and \Sfb
do not change and their uncertainties decrease to \num{0.020}.
This is considered as the statistical uncertainty for both \Sf and \Sfb.
The statistical correlation between $S_{f}$ and $S_{\bar f}$ is 60\%. This correlation is introduced by the flavour tagging and
by the production asymmetry. The distribution of the decay time with the overlaid projection
of the fit is shown in Fig.~\ref{fig:decaytime}.

\begin{figure}[tb]
  \includegraphics[width=0.45\textwidth]{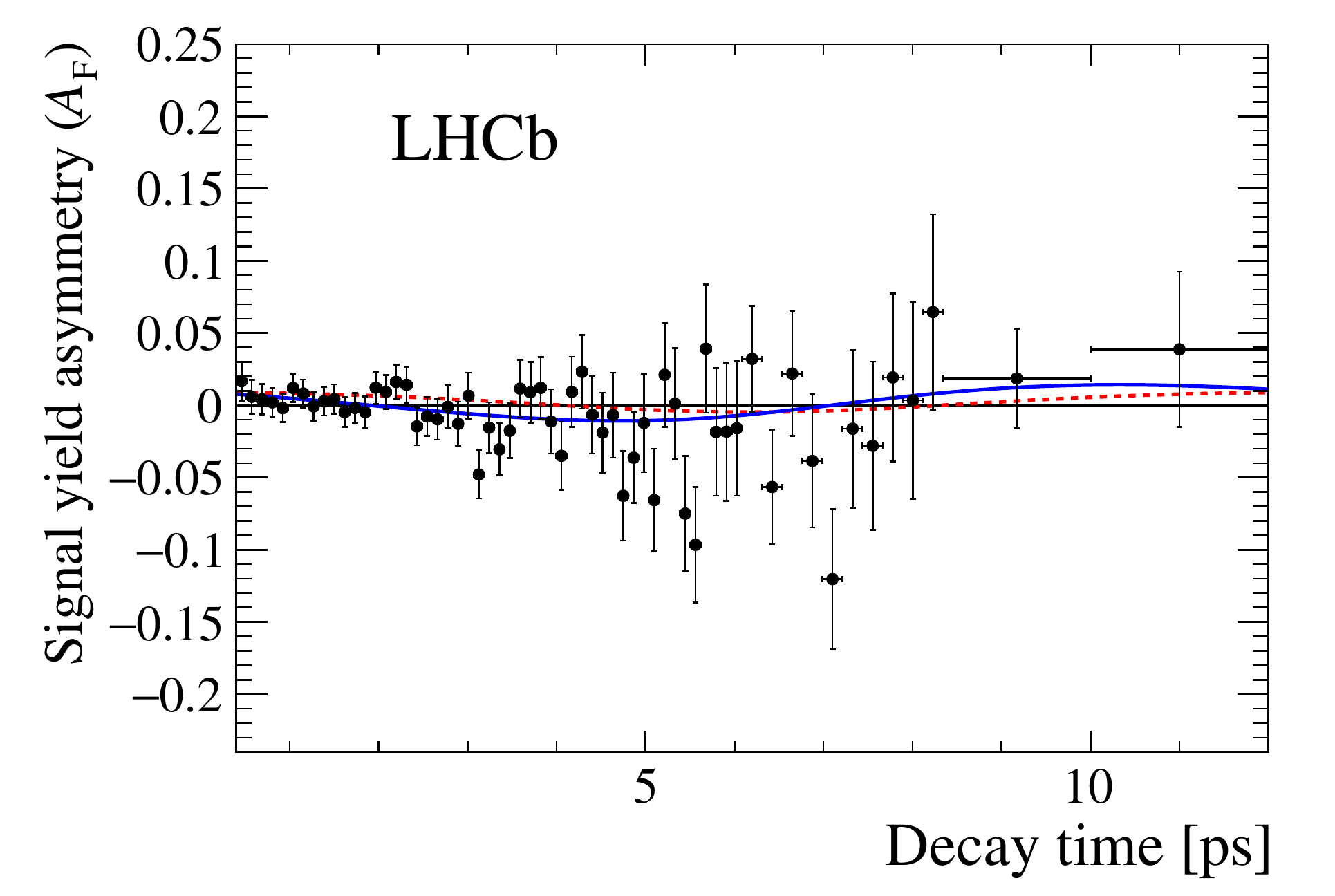}
  \includegraphics[width=0.45\textwidth]{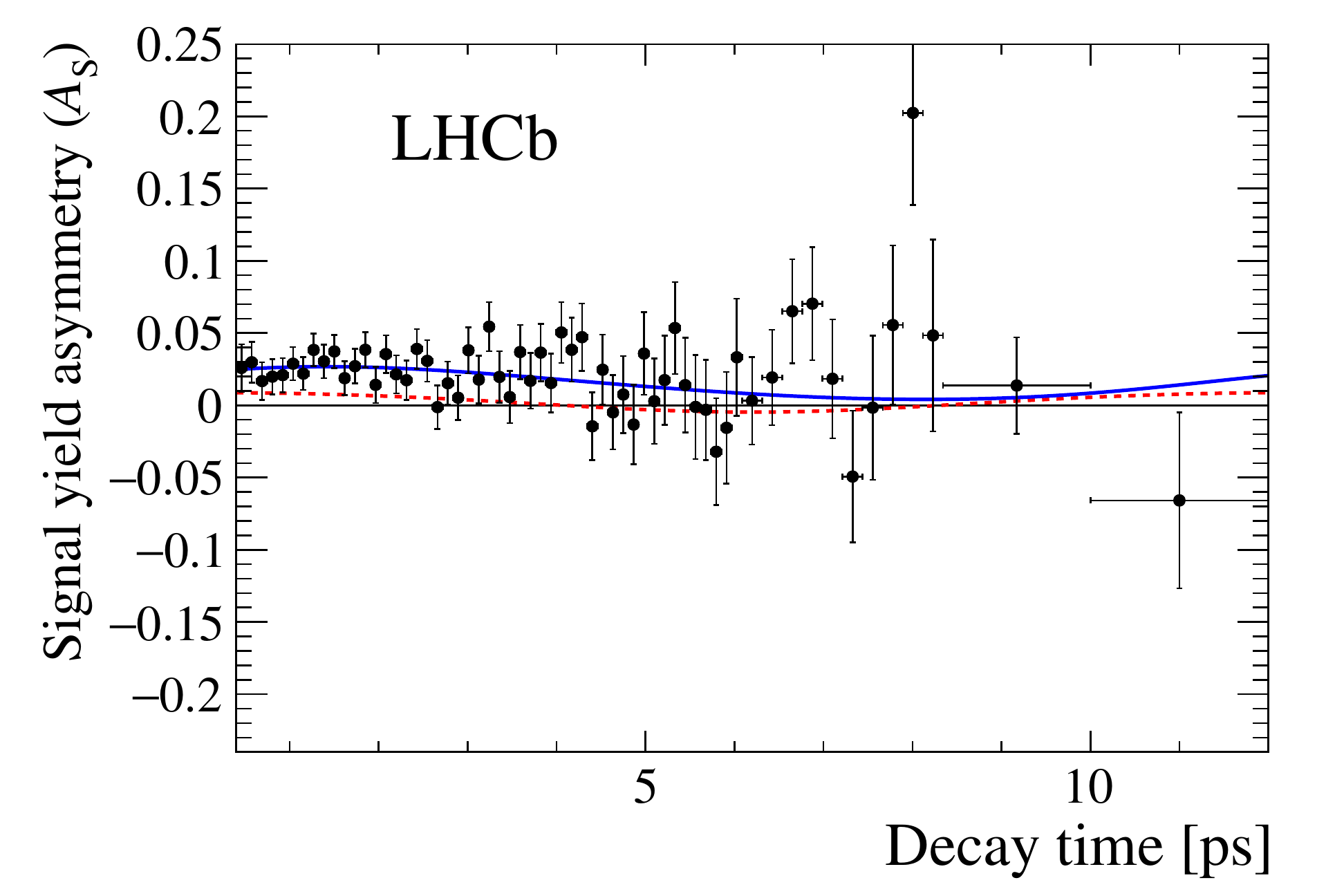}
\caption{\label{fig:asymmetries} Decay-time-dependent signal-yield asymmetries
for (left)  the favoured and (right) the suppressed decays.
The signal-yield asymmetries are defined in Eq.~\eqref{eq:CFAsym} and
Eq.~\eqref{eq:CSAsym}. The blue solid curve is the projection of the signal
PDF, the red dotted curve indicates the projection of the fit
when \CP conservation is imposed.}
\end{figure}

The values reported for \Sf and \Sfb result in a significance of $2.7\sigma$ for the \CP-violation hypothesis,
according to Wilks' theorem. Figure~\ref{fig:asymmetries} reports the decay-time-dependent
signal-yield asymmetries between candidates tagged as \Bz and \Bzb,
for the decays split according to the favoured (F) $\bar{b} \to \bar{c} u \bar{d}$ and the suppressed (S) $\bar{b} \to \bar{u} c \bar{d}$
transitions
\begin{align}
A_\text{F}&=\frac{\Gamma_{\Bd \to f} (t) - \Gamma_{\Bdb \to \bar{f}} (t)}{\Gamma_{\Bd \to f} (t) + \Gamma_{\Bdb \to \bar{f}} (t)}\label{eq:CFAsym}\\
A_\text{S}&=\frac{\Gamma_{\Bdb \to f} (t) - \Gamma_{\Bd \to \bar{f}} (t)}{\Gamma_{\Bdb \to f} (t) + \Gamma_{\Bd \to \bar{f}} (t)}\label{eq:CSAsym}.
\end{align}
The fit projections are overlaid to the asymmetries of the data, along with the curves expected when $\Sfb =-\Sf$ is imposed, \ie in the hypothesis of no \CP violation.

Several consistency checks are made by performing the fit on subsets of the data sample
split according to different data-taking conditions, tagging algorithms, number of tracks
in the event, and trigger requirements. These fits show good agreement with the result
presented here. The stability of the result is also analysed in bins of the transverse momentum
of the \Bz meson and in bins of the difference of pseudorapidity between the \Dm candidate and the companion pion.

The production asymmetry and the detection asymmetry are compared with results of independent LHCb measurements.
The values found in this analysis are \mbox{$A_{\rm{P}}= (-0.64\pm 0.28)\%$} and $A_{\rm{D}}= (0.86\pm0.19)\%$, where the uncertainties are statistical,
in agreement with those derived from Ref.~\cite{LHCb-PAPER-2016-062}, when accounting for the different kinematics of the signals.

The values of the flavour-tagging  parameters are also determined in control samples.
The $\Bu\to \Dzb\pip$ decay is used for the OS tagger. As the quarks
that accompany the $\bquark$ quark in $\Bu$ and $\Bd$ mesons differ,
the SS calibration function is studied with $\Bd\to\jpsi\Kstarz$ decays from a sample
that is disjoint to that used in the training of the BDT classifiers. In both
cases, distributions of $\pt$ and pseudorapidity of the $\Bd$ candidate,
number of tracks and PVs in the event, and the composition of software trigger decisions
are weighted to match those of the $\BdDpi$ signal sample. In the case of the
$\Bu\to \Dzb\pip$ mode, the decay-time distribution of the \Bu and \Dz mesons
are also weighted to match those of the \Bz and \Dm mesons of the signal decays,
while in the case of the $\Bd\to\jpsi\Kstarz$ decay the azimuthal angle of the \Bz is weighted
to match that of the $\BdDpi$ signal sample.
The charged pion produced in $\Bu \to \Dzb\pip$ decays directly identifies the \Bu flavour at production.
Therefore, the calibration of the OS tagger is achieved by counting the
number of correctly and incorrectly tagged signal candidates.
In contrast, the SS tagger calibration with $\Bd\to\jpsi\Kstarz$ decays requires the $\Bd$--$\Bdb$ flavour oscillations
to be resolved by using the decay time as an additional observable, since the amplitude
of the observed oscillation is related to the mistag fraction~\cite{LHCb-PAPER-2016-039}.
The values of the calibration parameters found in the control decays are in agreement
with those determined in the fit to the signal, with the largest deviation being of \num{2} standard deviations
for two of the $\Delta p_i$ parameters.


\section{Systematic uncertainties}
\label{sec:systematics}

Systematic uncertainties due to external measurements used in the fit are accounted for through
Gaussian constraints in the likelihood function. These parameters are the mixing frequency, $\dm$,
and the \Bz decay width, $\Gamma$. In order to disentangle these contributions from the statistical
uncertainty of $S_{f}$ and $S_{\bar f}$, the fit is repeated by fixing $\dm$ and $\Gamma$ to the
central values used in the constraints.
The systematic uncertainty due to the constraint on $\Gamma$ is found to be negligible, and that due to \dm
is \num{0.0073} and \num{0.0061} for \Sf and \Sfb, respectively. These are the largest systematic uncertainties
of \Sf and \Sfb and are found to be fully anticorrelated. The correlation of \dm with \Sf is
 $-34\%$ and that with \Sfb is $29\%$.

Validation of the entire analysis using ensembles of simulated signal candidates shows that the values
of $S_{f}$ and $S_{\bar f}$ are biased up to \num{0.0068} and \num{0.0018}, respectively.
The size of these potential biases are small and so are taken as a systematic uncertainty. The correlation
of these systematic uncertainties is $40\%$.

Variation of the fit to the $\Dm\pip$ invariant-mass distribution used to calculate the \emph{sWeights}
for the background subtraction leads to systematic uncertainties on $S_{f}$ and $S_{\bar f}$ of
\num{0.0042} and \num{0.0023}, respectively. Their correlation is 70\%.

The remaining systematic uncertainties are much smaller than those reported above. Hence,
the correlation between the  systematic uncertainty of \Sf and \Sfb for the sources that follow are neglected.
The systematic uncertainties associated with the PID efficiencies used in the fit to the $\Dm\pip$
invariant mass are also propagated by means of Gaussian constraints. These uncertainties take into
account the size of the calibration samples and the dependence of the results on the binning scheme
adopted for weighting the kinematic distributions of the particles of the control decays to match
those of the companion tracks. They contribute an uncertainty of \num{0.0008} to both $S_{f}$ and
$S_{\bar f}$.

The other sources of systematic uncertainty are calculated by means of pseudoexperiments,
where samples of the same size as the data are generated by sampling the PDF with parameters fixed to the value found in data.
In the generation of the pseudoexperiments the PDF is modified to consider alternative
models according to the source of systematic uncertainty under investigation. The generated sample is
then fit with the nominal model. For each parameter, the mean of the distribution of the residuals is
considered, $(S_i^{\textrm{gen}} - S_i^{\textrm{fit}})$, from \num{1000} pseudoexperiments as the systematic
uncertainty. If the mean differs from zero by less than one standard deviation, the error on the mean is taken
as the systematic uncertainty.

To test the impact of the choice of the calibration models, pseudoexperiments are generated using for
the SS calibration the nominal model, while for the OS the degree of the
polynomial used in the model is reduced by one unit compared to the nominal model. In the fit for both
taggers the degrees of the calibration models are increased by one degree compared to that used to generate
the pseudoexperiments. The systematic uncertainties are determined to be \num{0.0008} and  \num{0.0016} for
$S_{f}$ and $S_{\bar f}$, respectively.

Assuming values for the flavour-tagging efficiency asymmetries different from zero, based on what is found in
simulation, leads to systematic uncertainties of \num{0.0012} and \num{0.0015} for $S_{f}$ and $S_{\bar f}$,
respectively.

A different decay-time acceptance model is used in generation by considering new boundaries of
the subranges of the spline functions. This results in a systematic uncertainty of \num{0.0007}
for both $S_{f}$ and $S_{\bar f}$.

Mismodelling of the decay-time resolution is also considered by increasing and decreasing the
nominal resolution by $20\fs$. The largest residuals are considered as the systematic uncertainties,
and are \num{0.0012} and \num{0.0008} for $S_{f}$ and $S_{\bar f}$, respectively.

A value for $C_f=-C_{\bar{f}}$ different from 1, based on the value of $r_{D\pi}$ from
Refs.~\cite{Aubert:2008zi,Das:2010be} is assumed, resulting in a variation of \num{0.0006}
for both \Sf and \Sfb. By assigning to \DG a value different from zero and equal to the world-average
value plus its uncertainty~\cite{HFLAV16} leads to a systematic uncertainty of \num{0.0007} on both
\Sf and \Sfb.

The sources of systematic uncertainties are summarised in Table~\ref{tab:summarySyst}. They total
\Sfsyst and \Sfbsyst for $S_{f}$ and $S_{\bar f}$, respectively, with a correlation of $-41\%$.

\begin{table}[htbp]
        \centering
        \caption{Systematic uncertainties on the \CP asymmetries $S_{f}$ and
        $S_{\bar f}$. The total uncertainty is the sum in quadrature of the individual contributions.}
        \begin{tabular}{lcc}
		\hline
		Source & $S_{f}$ & $S_{\bar f}$ \\
		\hline
		uncertainty of \dm & \num{0.0073} & \num{0.0061} \\
		fit biases & \num{0.0068} & \num{0.0018} \\
		background subtraction  & \num{0.0042} & \num{0.0023} \\
        PID efficiencies & \num{0.0008} & \num{0.0008} \\
        flavour-tagging models & \num{0.0011} & \num{0.0015} \\
        flavour-tagging efficiency asymmetries & \num{0.0012} & \num{0.0015} \\
        $\epsilon(t)$ model & \num{0.0007} & \num{0.0007} \\
        assumption on \DG & \num{0.0007} & \num{0.0007} \\
        decay-time resolution  & \num{0.0012} & \num{0.0008} \\
        assumption on $C$ & \num{0.0006} & \num{0.0006} \\
	\hline
	total & 0.0111 & 0.0073 \\
        \hline
        statistical uncertainty & 0.0198 & 0.0199 \\
	\hline
	\end{tabular}
        \label{tab:summarySyst}
\end{table}


\section{\boldmath Interpretation of the $\CP$ asymmetries}
\label{sec:interpretation}

The values of $S_{f}$ and $S_{\bar f}$ are interpreted in terms of the angle
$2\beta+\gamma$, the ratio of amplitudes $r_{D\pi}$, and the strong phase $\delta$, using the
statistical method described in Ref.~\cite{LHCb-PAPER-2016-032}.

By taking external measurements of $r_{D\pi}$, confidence intervals for $|\sin(2\beta+\gamma)|$ and $\delta$ are derived.
The ratio $r_{D\pi}$ is calculated from the branching fraction of $\Bd \to \Dsp \pim$ decays, assuming
SU(3) symmetry, following the same relation used in Refs.~\cite{Das:2010be,Aubert:2008zi}:
\begin{equation}
r_{D\pi}= \tan \theta_c  \frac{f_{D^+}}{f_{D_s}}\sqrt{\frac{\mathcal{B}(\Bd \to \Dsp \pim)}{\mathcal{B}(\Bd \to \Dm \pip)}},
\end{equation}
where $\tan\theta_c=0.23101\pm0.00032$ is the tangent of the Cabibbo angle from Ref.~\cite{CKMfitter2005},
\mbox{$f_{D_s}/f_{D^+}=1.173\pm0.003$} is the ratio of decay constants~\cite{Aoki:2016frl,Bazavov:2014wgs,Carrasco:2014poa}, and
\mbox{$\mathcal{B}(\Bd\to\Dsp\pim)=(2.16\pm0.26)\times10^{-5}$} and \mbox{$\mathcal{B}(\Bd\to\Dm\pip)=(2.52\pm0.13)\times10^{-3}$}
are branching fractions taken from Ref.~\cite{PDG2017}. We determine \mbox{$r_{D\pi}=0.0182\pm0.0012\pm0.0036$},
where the second uncertainty accounts for possible nonfactorizable SU(3)-breaking effects, considered to be 20\% of
the value of $r_{D\pi}$ as suggested in Ref.~\cite{DeBruyn:2012jp}. In addition, using the known value of $\beta = (22.2 \pm 0.7)^\circ$~\cite{HFLAV16},
confidence intervals for $\gamma$ are determined.

\begin{figure}[tbp]
  \centering
        \includegraphics[width=0.5\textwidth]{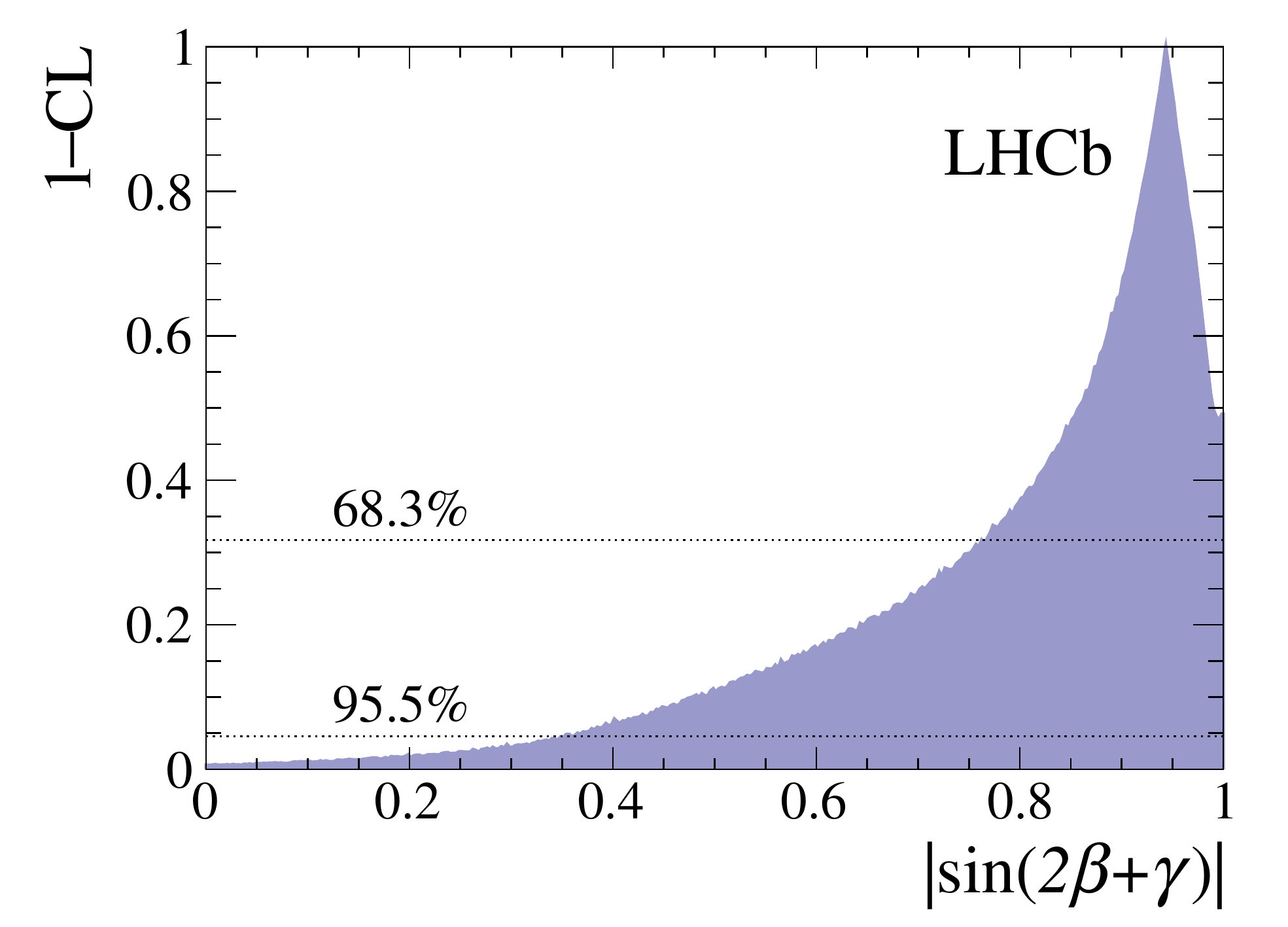}
        \caption{1--CL as a function of $|\sin(2\beta+\gamma)|$.}
        \label{fig:gammacombo1}
\end{figure}

\begin{figure}[htb]
	\includegraphics[width=0.45\textwidth]{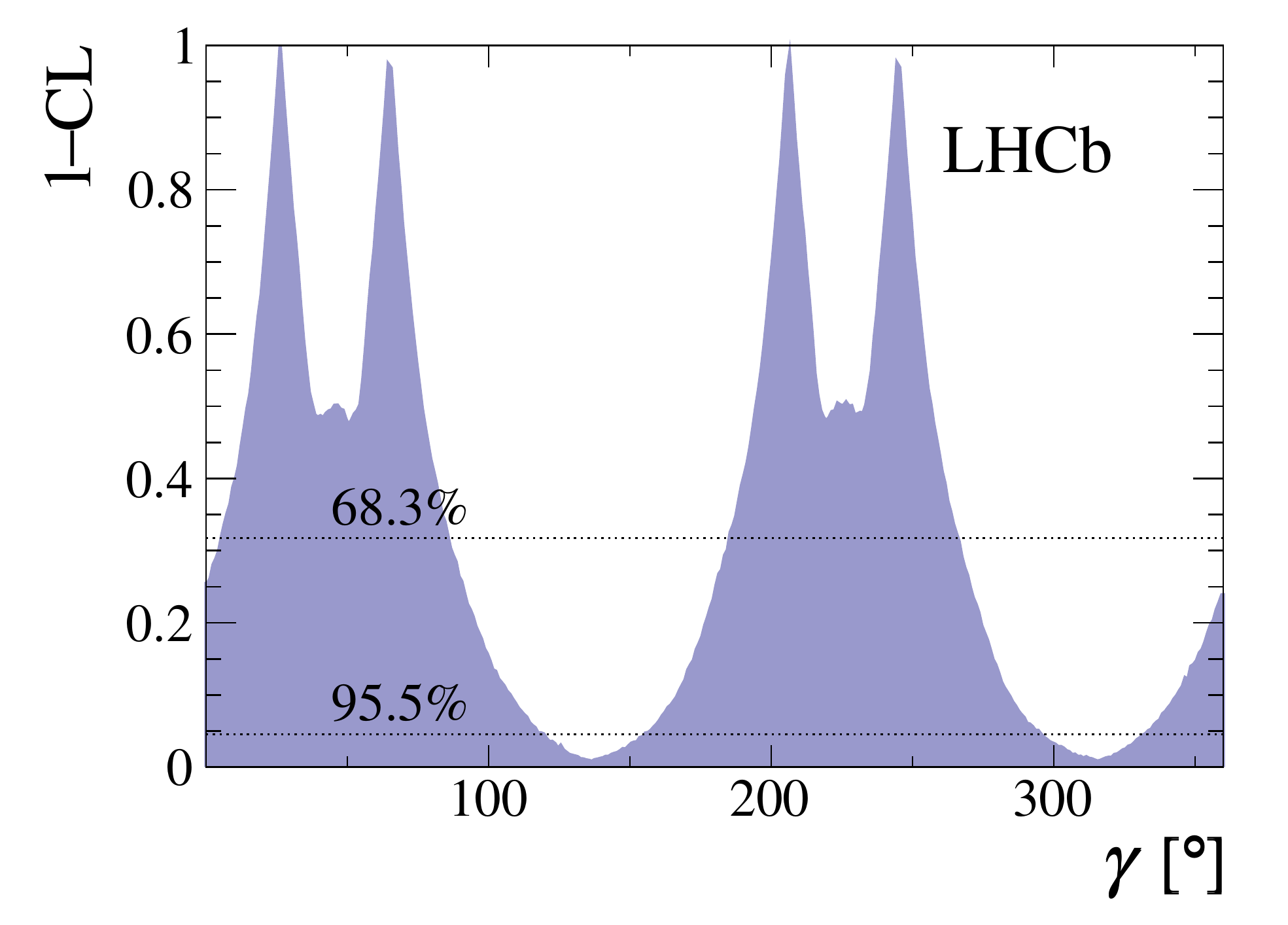}
	\includegraphics[width=0.45\textwidth]{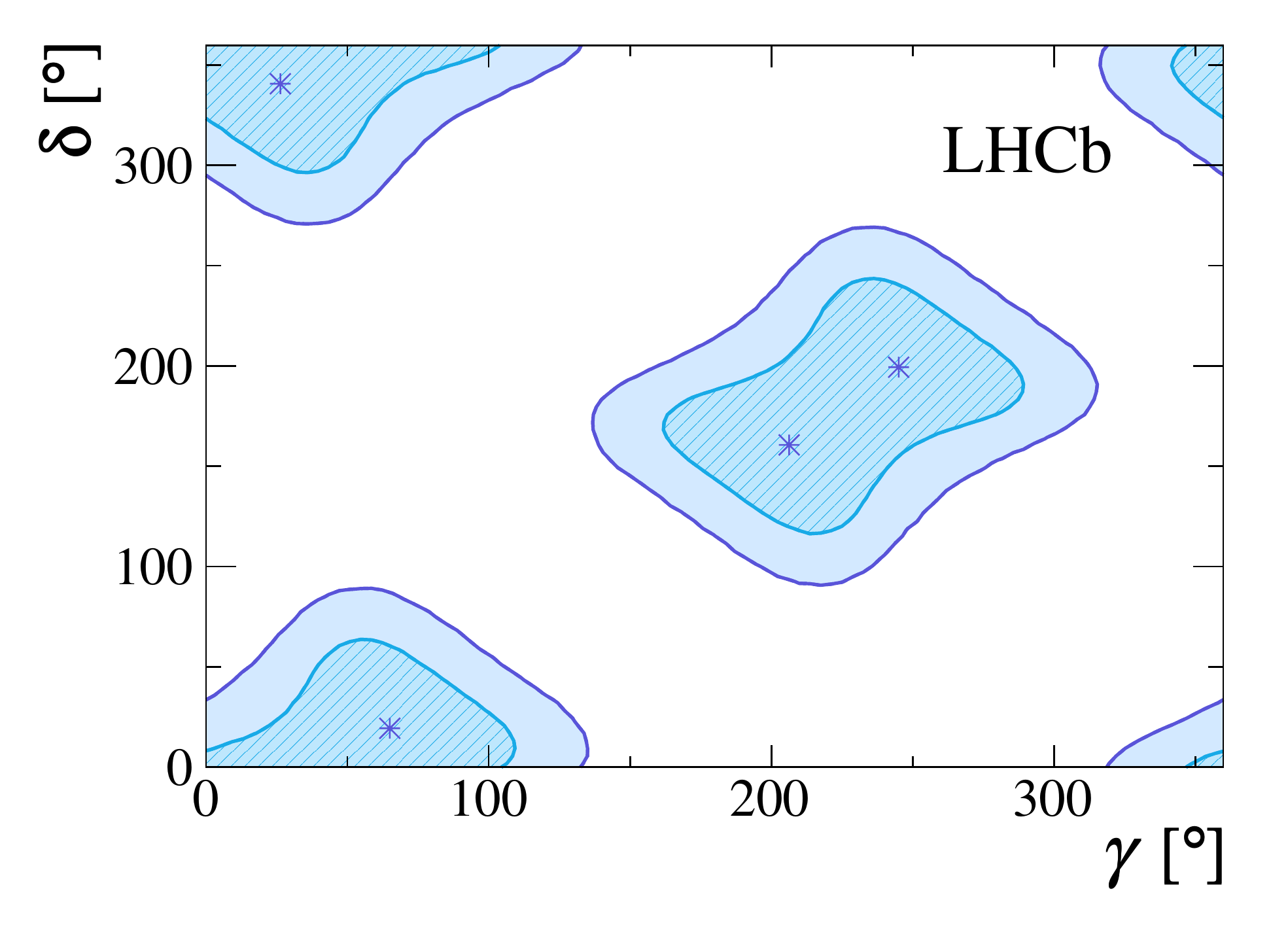}
	\caption{(Left) 1--CL as a function of  $\gamma$ and (right) confidence regions for $\gamma$ and
	$\delta$. The confidence regions hold the $39\%$ and $87\%$ CL. Points denote the preferred values.}
    \label{fig:gammacombo2}
\end{figure}

The confidence intervals are
\begin{align*}
|\sin(2\beta+\gamma)| \in \magSinTwoBplusGCL\,, \\
\gamma \in \gammaCL\,,\\
\delta \in \deltaCL\,,
\end{align*}
 all at the 68\% confidence level (CL). The uncertainties on $r_{D\pi}$ and $\beta$ have a negligible impact on these values.
 The intervals are illustrated in Figs.~\ref{fig:gammacombo1} and \ref{fig:gammacombo2}.

\section{Conclusion}
\label{sec:conclusion}

A measurement of the $\CP$ asymmetries $\Sf$ and $\Sfb$ in the decay
$\BdDpi$  is reported. The decay candidates are reconstructed in a data set collected with the \lhcb experiment at centre-of-mass energies
of $7$ and $8\tev$, corresponding to an integrated luminosity of $3.0\invfb$. We measure
\begin{align*}
  \Sf &= \Sfval \pm \Sfstat\stat \pm \Sfsyst \syst, \\
  \Sfb &= \Sfbval \pm \Sfbstat\stat \pm \Sfbsyst \syst,
\end{align*}
with a correlation of $60\%$ ($-41\%$) between the statistical (systematic) uncertainties.
These values are in agreement with, and more precise than, measurements from the Belle and BaBar collaborations~\cite{Aubert:2006tw,PhysRevD.73.092003}.
This measurement, in combination with the external inputs of $r_{D\pi}$ and $\beta$,
constrains the CKM angle $\gamma$ to be in the interval \gammaCL at the 68\% confidence level.


\section*{Acknowledgements}
%
%
\noindent We express our gratitude to our colleagues in the CERN
accelerator departments for the excellent performance of the LHC. We
thank the technical and administrative staff at the LHCb
institutes. We acknowledge support from CERN and from the national
agencies: CAPES, CNPq, FAPERJ and FINEP (Brazil); MOST and NSFC
(China); CNRS/IN2P3 (France); BMBF, DFG and MPG (Germany); INFN
(Italy); NWO (The Netherlands); MNiSW and NCN (Poland); MEN/IFA
(Romania); MinES and FASO (Russia); MinECo (Spain); SNSF and SER
(Switzerland); NASU (Ukraine); STFC (United Kingdom); NSF (USA).  We
acknowledge the computing resources that are provided by CERN, IN2P3
(France), KIT and DESY (Germany), INFN (Italy), SURF (The
Netherlands), PIC (Spain), GridPP (United Kingdom), RRCKI and Yandex
LLC (Russia), CSCS (Switzerland), IFIN-HH (Romania), CBPF (Brazil),
PL-GRID (Poland) and OSC (USA). We are indebted to the communities
behind the multiple open-source software packages on which we depend.
Individual groups or members have received support from AvH Foundation
(Germany), EPLANET, Marie Sk\l{}odowska-Curie Actions and ERC
(European Union), ANR, Labex P2IO and OCEVU, and R\'{e}gion
Auvergne-Rh\^{o}ne-Alpes (France), Key Research Program of Frontier
Sciences of CAS, CAS PIFI, and the Thousand Talents Program (China),
RFBR, RSF and Yandex LLC (Russia), GVA, XuntaGal and GENCAT (Spain),
Herchel Smith Fund, the Royal Society, the English-Speaking Union and
the Leverhulme Trust (United Kingdom).




\addcontentsline{toc}{section}{References}
\setboolean{inbibliography}{true}
\bibliographystyle{LHCb}
\bibliography{main,LHCb-PAPER,LHCb-CONF,LHCb-DP,LHCb-TDR}

\newpage



\newpage
\centerline{\large\bf LHCb collaboration}
\begin{flushleft}
\small
R.~Aaij$^{43}$,
B.~Adeva$^{39}$,
M.~Adinolfi$^{48}$,
Z.~Ajaltouni$^{5}$,
S.~Akar$^{59}$,
P.~Albicocco$^{18}$,
J.~Albrecht$^{10}$,
F.~Alessio$^{40}$,
M.~Alexander$^{53}$,
A.~Alfonso~Albero$^{38}$
% ...

{\footnotesize \it
% ...
$ ^{5}$Clermont Universit{\'e}, Universit{\'e} Blaise Pascal, CNRS/IN2P3, LPC, Clermont-Ferrand, France\\
% ...
$ ^{10}$Fakult{\"a}t Physik, Technische Universit{\"a}t Dortmund, Dortmund, Germany\\
% ...
$ ^{18}$INFN Laboratori Nazionali di Frascati, Frascati, Italy\\
% ...
$ ^{38}$ICCUB, Universitat de Barcelona, Barcelona, Spain\\
$ ^{39}$Instituto Galego de F{\'\i}sica de Altas Enerx{\'\i}as (IGFAE), Universidade de Santiago de Compostela, Santiago de Compostela, Spain\\
$ ^{40}$European Organization for Nuclear Research (CERN), Geneva, Switzerland\\
% ...
$ ^{43}$Nikhef National Institute for Subatomic Physics, Amsterdam, The Netherlands\\
% ...
$ ^{48}$H.H. Wills Physics Laboratory, University of Bristol, Bristol, United Kingdom\\
% ...
$ ^{53}$School of Physics and Astronomy, University of Glasgow, Glasgow, United Kingdom\\
% ...
$ ^{59}$University of Cincinnati, Cincinnati, OH, United States\\
% ...
}
\end{flushleft}



\end{document}